\documentclass[letterpaper, 10pt, conference]{ieeeconf}  

\IEEEoverridecommandlockouts                              

\usepackage{graphicx}
\usepackage{url}  
\usepackage{amsmath}
\usepackage{todonotes}
\usepackage{multirow}

\begin{document}

\title{\LARGE \bf
    Robust, General, and Low Complexity Acoustic Scene \\ Classification Systems and An Effective Visualization \\ for Presenting a Sound Scene Context
}

\author{Lam~Pham,  
        Dusan~Salovic,             
        Anahid~Jalali, 
        Alexander~Schindler,
        Khoa~Tran,
        Canh~Vu,
        Phu~X. Nguyen
	\thanks{L. Pham, A. Jalali, and A. Schindler are with Center for Digital Safety \& Security, Austrian Institute of Technology (AIT), Austria.   
D. Salovic is with University of Technology, Austria.
P. X. Nguyen is with University of FPT, Vietnam. 
K. Tran is with Da Nang University, Vietnam. 
C. Vu is with University of Technology of Compiègne, France.}%
}

\maketitle
\thispagestyle{empty}
\pagestyle{empty}

\begin{abstract}
In this paper, we present a comprehensive analysis of Acoustic Scene Classification (ASC), the task of identifying the scene of an audio recording from its acoustic signature.
In particular, we firstly propose an inception-based and low-footprint ASC model, referred to as the ASC baseline.
The proposed ASC baseline is then compared with benchmark and high-complexity network architectures of MobileNetV1, MobileNetV2, VGG16, VGG19, ResNet50V2, ResNet152V2, DenseNet121, DenseNet201, and Xception.
Next, we improve the ASC baseline by proposing a novel deep neural network architecture which leverages residual-inception architectures and multiple kernels.
Given the novel residual-inception (NRI) model, we further evaluate the trade off between the model complexity and the model accuracy performance.
Finally, we evaluate whether sound events occurring in a sound scene recording can help to improve ASC accuracy, then indicate how a sound scene context is well presented by combining both sound scene and sound event information.
We conduct extensive experiments on various ASC datasets, including Crowded Scenes, IEEE AASP Challenge on Detection and Classification of Acoustic Scenes and Events (DCASE) 2018 Task 1A and 1B, 2019 Task 1A and 1B, 2020 Task 1A, 2021 Task 1A, 2022 Task 1.
The experimental results on several different ASC challenges highlight two main achievements; the first is to propose robust, general, and low complexity ASC systems which are suitable for real-life applications on a wide range of edge devices and mobiles; the second is to propose an effective visualization method for comprehensively presenting a sound scene context. 


\end{abstract}

\section{Introduction}
\label{intro}

Acoustic Scene Classification (ASC), one of two main tasks of the machine hearing research~\cite{lyon2017_bk}, aims at detecting surrounding environments such as in a bus, in an shopping mall, or on a street. 
By detecting the current sound scene context, edge devices could make use of this useful information to enable them to respond appropriately or adjust certain functions, then opening up various applications: to integrate an ASC component into a robotic system~\cite{brian1998_ap01}, a mobile application~\cite{maleh1999_ap01}, or a sensor system~\cite{jakob2017_acoustic} as one of main functions; to support sound event detection when these sound events are mixed in real-world environments~\cite{toni2013_ap01}. 
Considering a general recording of an acoustic environment, it contains not only a background sound field but also various foreground events.  
If the background is considered as the noise and the foreground is referred to as the signal, it can be seen that the signal-to-noise ratio exhibits high variability due to the diverse range of environments or recording conditions.
To further complicate matters, if a sound event can occur in a long time, it could be considered as background in certain contexts.
For example, an audio recording on \textit{pedestrian street} may present a quiet background, but the sound of the \textit{engine} of traffic passes is considered as the foreground events.
However, a lengthy \textit{engine} sound in an \textit{on train/bus} recording would be considered a background sound.
Furthermore, both background and foreground contain true noise -- continuous, periodic or aperiodic acoustic signals that interfere with the understanding of the scene.
Recently, two new issues of mismatched recording devices~\cite{task1b_2018, task1_2020_1ads} and low-complexity model~\cite{task1a_2021, task1a_2022} within ASC task have been indicated.
In particular, the first issue of mismatched recording devices causes very different distribution of energy across the frequency dimensions of audio spectrograms~\cite{missmatch_cite}, which leads classification models to misclassify.
Meanwhile, the second issue of high-complexity ASC models prevents to integrate these models in edge devices or mobiles with memory limitation.
All these challenges mentioned make acoustic scene classification (ASC) task particularly challenging.

To deal with the challenges mentioned above, recent ASC papers have tended to focus on two main approaches. 
The first aims at solving the lack of discriminative information by exploiting various methods of low-level feature extraction.
In particular, an input audio is transformed into different two-dimensional spectrogram representations.
Then, these spectrograms are independently trained with back-end deep learning models.
Finally, independent models' results are fused to achieve the best performance.
For instances, log-Mel spectrogram was combined with constant-Q transform (CQT)~\cite{hos_dca_18}, gammatonegram~\cite{huy_j01}, or draw audio~\cite{mel_audio_2021}.
To evaluate a wavelet-transform derived spectrogram representation, Ren et al.~\cite{dc_17_jou_64} compared results from STFT spectrograms and a combination of \textit{Bump} and \textit{Morse} scalograms.
By exploiting channel information, Sakashita and Aono~\cite{yuma_dca_18} generated multi-spectrogram inputs from two channels, the average and side channels, and even explored separated harmonic and percussive spectrograms from mono channels.
The approach of using multiple spectrograms has proven powerful to tackle the issue of mismatched recoding devices.
Indeed, a combination of log-Mel and Mel-based nearest neighbor filter (NNF) spectrograms in ~\cite{truc_dca_18} helps to achieve the top-1 on DCASE 2018 Task 1B blind Test set and the top-4 on DCASE 2018 Task 1B Development set.
Meanwhile, the authors in~\cite{mul_spec_2020} conducted various ensemble methods on log-Mel, gammatonegram, CQT, and MFCC spectrograms, then achieved the top-6 on DCASE 2020 Task 1A blind Test set and the top-1 on DCASE 2020 Task 1A Development set.
Although this approach shows effective to deal with the issue of mismatched recording devices, it presents the issue of large footprint model as using ensemble of multiple individual classifiers.

Instead of using multiple spectrogram inputs, the second approach tends to deploy more complex deep learning architectures, especially focusing on exploring the frequency bands of audio spectrograms.
For instances, authors in~\cite{mul_sub_spec_01} split the entire log-Mel spectrograms into three sub spectrograms across the frequency dimension. 
Then, each sub spectrogram was learned by a ResNet-based network architecture before concatenating together. 
Meanwhile, Phaye et al.~\cite{phaye_dca_18} proposed a SubSpectralNet network which comprises multiple sub-networks with parallel branches to extract discriminative information from 30 sub log-Mel spectrograms. 
Focusing on the frequency normalization, authors in~\cite{freq_norm_01} proposed a novel Residual Normalization method and a residual-based network architecture, which showed effective to improve the ASC performance and achieved the top-1 on DCASE 2021 Task 1A blind Test set and the top-4 on DCASE 2021 Task 1A Development set.
However, to achieve the best performance, some papers from the second approach have still applied ensemble methods of multiple models~\cite{ens_prun_01, ens_prun_03, quan_ens_05, ens_06}, which increases the model complexity.

To deal with the issue of large footprint models as using complex network architectures, ensemble of multiple models, or ensemble of multiple spectrogram inputs, pruning~\cite{ens_prun_01, quan_prun_02, ens_prun_03, prun_04} and quantization~\cite{quan_prun_02, quan_ens_05} techniques have been widely applied.
While quantization techniques feasibly help the model reduce to 1/4 of the original size (i.e, 32 bit with floating point format presenting for 1 trainable parameter is quantized to 8 bit with integer format~\cite{quantization_google}), pruning techniques prove that models can be reduced to 1/10 of the original sizes~\cite{quan_prun_02, prun_04}.

Looking at the recent approaches surveyed above, we can see that: \textbf{(I)} While ensembles of multiple spectrograms or complex network architectures can help to enhance the ASC performance as well as effectively to deal with the mismatched recording devices, these approaches present large footprint models which are not suitable for applications on edge devices or mobiles.
However, it is fact that none of research has provided an analysis of the trade off between the performance and the complexity of an ASC model.
Additionally, almost proposed low-complexity ASC models currently leverage pruning techniques.
However, although pruning techniques prove to reduce the model complexity significantly, the pruning parameters are not removed from the proposed network architecture and they still occupy the memory of edge devices as well as cost computation same as the none-pruning parameters.
Therefore, the recent DCASE 2022 Task 1 challenge~\cite{task1a_2022}, which focuses on the issue of low-complexity ASC model, requires not to use pruning techniques. 
\textbf{(II)} As a sound scene can contain different types of sound events and some sound events are distinct for certain sound scene, this inspires that exploring sound event information in a sound scene recording can help to further improve the ASC performance.
However, just a few of researches~\cite{scene_by_event_01, jung_asc_aed} leveraged sound event information for enhancing ASC systems and none of research has deeply analyzed the relationship between sound scene and sound events.
In this paper, we therefore fill these two gaps of the ASC research, then mainly contribute:
\begin{enumerate}
\item Firstly, we propose an inception based ASC baseline framework. We then evaluate and compare the proposed ASC baseline with benchmark and high-complexity network architectures (e.g., MobileNet, VGG, DenseNet, etc.). Our experimental results indicate that shallow and wider neural networks are more effective rather than deeper architectures with a trunk of convolutional layers for the ASC task.\\

\item Secondly, we propose a novel residual-inception (NRI) based neural network to improve the ASC baseline. Our proposed network architecture and ensemble of multiple spectrogram inputs prove effective to deal with the issue of mismatched recording devices as well as are competitive state-of-the-art ASC systems on various datasets. \\

\item Thirdly, we successfully apply multiple model compression techniques: channel reduction (CR), channel deconvolution (CD), and quantization (Qu) to achieve low-complexity ASC systems but still perform robust. 
We achieve two low complexity models: the first model with less than 20 MB memory occupation which satisfies a wide range of edge devices and mobiles surveyed in~\cite{model_size_sur_01, model_size_sur_02}, and the second model with less than 128 KB memory occupation which is suitable for very limited-memory embedded systems such as STM32L496@80MHz or Arduino Nano 33@64MHz.
Notably, one trainable parameter is presented by 32 bit with floating format in our experiments in this paper.
Given low-complexity ASC models, we provide a comprehensive analysis of the trade off between the performance and the complexity of proposed ASC systems. \\

\item We finally evaluate the role of sound events in a sound scene recording, indicate how sound events can help to improve the ASC performance. 
Given the analysis of sound events, we provide an effective visualization method to present a sound scene context by combining both sound scene and sound event information.

\end{enumerate}

Rather than selecting a single task, we evaluate over a wide range of datasets of: 
Crowded-Scene~\cite{dataset_crowded_scene}, DCASE 2018 Task 1A and 1B~\cite{dc_2018_dataset}, DCASE 2019 Task 1A and 1B~\cite{dc_2019_dataset}, DCASE 2020 Task1A~\cite{dataset_dcase_2021_2022}, DCASE 2021 Task 1A~\cite{dataset_dcase_2021_2022}, and DCASE 2022 Tasks 1~\cite{dataset_dcase_2021_2022}. 
We will see that the performance of our proposed system is competitive with the state-of-the-art systems.

\section{Evaluating datasets}
\label{data_set}

As each following section in this paper not only describes ASC systems but also evaluate and discusses experimental results, we firstly present datasets of sound scenes used in this paper and indicate why and which datasets are evaluated in certain sections. 

\textbf{Crowded Scenes (Cr-Sc)~\cite{dataset_crowded_scene}:} contains 341 videos collected from YouTube (in-the-wild scenes), which presents a total recording time of nearly 29.06 hours.
These videos were then split into 10-second video segments, each of which was annotated by one of five categories: `Riot', `Noise-Street', `Firework-Event', `Music-Event', or `Sport-Atmosphere'. 
Notably, 10-second video segments split from an original video are not presented in both Train and Test subsets to make the data distribution different between these two subsets. 
In this paper, we extract audio recordings from these video segments and follow the splitting method as proposed in~\cite{dataset_crowded_scene} to evaluate this dataset. 


\textbf{DCASE 2018 Task 1A~\cite{dc_2018_dataset} and DCASE 2019 Task 1A~\cite{dc_2019_dataset} Development sets:} DCASE 2018 Task 1A Development set (DC-18-1A) comprises 8640 10-second segments with a total recording time of 24 hours. 
The dataset was recorded from one device, referred to as the device A.
DCASE 2019 Task 1A Development set (DC-19-1A) reused all DCASE 2018 Task 1A Development set and more data was added (i.e., Additional audio recordings were also recorded on the same device A), then create 14400 10-second segments with 40 hours recording time.
As audio recordings from these both datasets are from one device A, they are used to evaluate ASC task regardless of the issue of mismatched recording devices. 
To evaluate these DCASE 2018 and 2019 Task 1A datasets, we follow DCASE challenges, then separated Development set into Training (i.e., 6122 and 9185 10-second segments for DCASE 2018 and 2019, respectively) and Evaluating (i.e., 2518 and 4185 10-second segments for DCASE 2018 and 2019, respectively) subsets for training and evaluating processes respectively.

\textbf{DCASE 2018 Task 1B~\cite{dc_2018_dataset} and DCASE 2019 Task 1B~\cite{dc_2019_dataset} Development sets:} DCASE 2018 Task 1B Development set (DC-18-1B) reused all DCASE 2018 Task 1A Development set recently mentioned, and added more data recorded from two other devices, referred to as the device B and the device C, with only 2 hours of recording time (72 10-second segments) for each device B or C.
Similarly, DCASE 2019 Task 1B Development set (DC-19-1B) also reused DCASE 2019 Task 1A Development, and 3 hours of recording time on each device B and C were further added. 
As audio recordings from these both datasets are from three different devices (A, B, and C) with a limitation of recording time on device B and C, they are used to evaluate ASC task concerning the issue of mismatched recording devices. 
To evaluate these datasets, we follow DCASE challenges, then separated Development sets into Training and Evaluating subsets for training and evaluating processes respectively.

\textbf{DCASE 2020 Task 1A Development set (DC-20-1A)~\cite{dataset_dcase_2021_2022}:} This is currently the largest dataset proposed for ASC task which comprises 23040 10-second segments with a total recording time of 64 hours. 
The dataset was recorded from three real devices namely A, B, and C with 40 hours, 3 hours, and 3 hours, respectively.
In particular, DCASE 2019 Task 1B Development set was completely reused in DCASE 2020 Task 1A Development.
Additionally, synthesized audio recordings namely from S1 to S6 with 3-hour recording time for each synthesized device were added. 
Notably, audio recordings from S4, S5, and S6 are not presented in Training subset to evaluate unseen samples.
As audio recordings are from both different real and synthesized devices, this dataset is proposed to evaluate ASC task with the issue of mismatched recording devices. 
To evaluate DCASE 2020 Task 1A Development dataset, we follow the challenge, then use two subsets of the DCASE 2020 Task 1A Development set (i.e. Not all audio recordings in this Development set is used), referred to as Training and Evaluating subsets, for training and evaluating processes, respectively.

This dataset is also used in \textbf{DCASE 2021 Task 1A (DC-21-1A) and DCASE 2022 Task 1 (DC-22-1)} challenges for evaluating both issues of mismatched recording devices and low-complexity models. 
While DCASE 2021 Task 1A challenge evaluates low-complexity models on 10-second segments, it is more challenging in DCASE 2022 Task 1 challenge as this task requires to evaluate on 1-second segments and does not allow to use pruning techniques.
The Training/Evaluating splitting methods used for DCASE 2021 Task 1A and DCASE 2022 Task 1 challenges are same as DCASE 2020 Task 1A challenge.

Given these updated and benchmark sound scene datasets mentioned above, we can see that DCASE 2020 Task 1A Development set includes all DCASE 2018/2019 Task 1A/1B Development sets and it is proposed to evaluated the issue of low complexity model in DCASE 2021 Task 1A and DCASE 2022 Task 1 challenges.
We, therefore, firstly use DCASE 2020 Task 1A Development dataset for: (1) evaluating our proposed ASC baseline in Section~\ref{baseline_section}, (2) evaluating our proposed novel residual-inception neural network architecture in Section~\ref{novel_section}, (3) analyzing the trade off between ASC system performance and complexity in Section~\ref{tradeoff_section}, and (4) evaluating how sound event information can help to improve ASC accuracy performance in Section~\ref{event_section}.
These work and the corresponding sections recently mentioned focus on how to achieve robust, general, and low-complexity ASC models.

Secondly, while all DCASE datasets present ten daily scenes, Crowded Scenes dataset was proposed to classify five very noise sound contexts.
This inspires us to define a new dataset of sound scene contexts which comprises both DCASE 2020 Task 1A and Crowded Scenes datatsets.
As the new dataset presents diverse scene contexts, a wide range of sound events can be observed.
Additionally, it is fact that specific sound events can only occur in certain sound scene such as the gun sound or explosion in a riot context or the loud music in a music event. 
We, therefore, make use of statistic information of sound events in a sound scene context as well as ASC systems proposed in previous sections to develop a visualization method.
The proposed visualization method helps to present a sound scene context more comprehensive in Section~\ref{visual_section}.

Finally, all datasets mentioned are evaluated with our proposed models and compared with the state-of-the-art systems in Section~\ref{sota_section}.

\section{Propose an ASC baseline system and compare to benchmark network architectures}
\label{baseline_section}
\begin{figure}[t]
    \centering
    \includegraphics[width =1.0\linewidth]{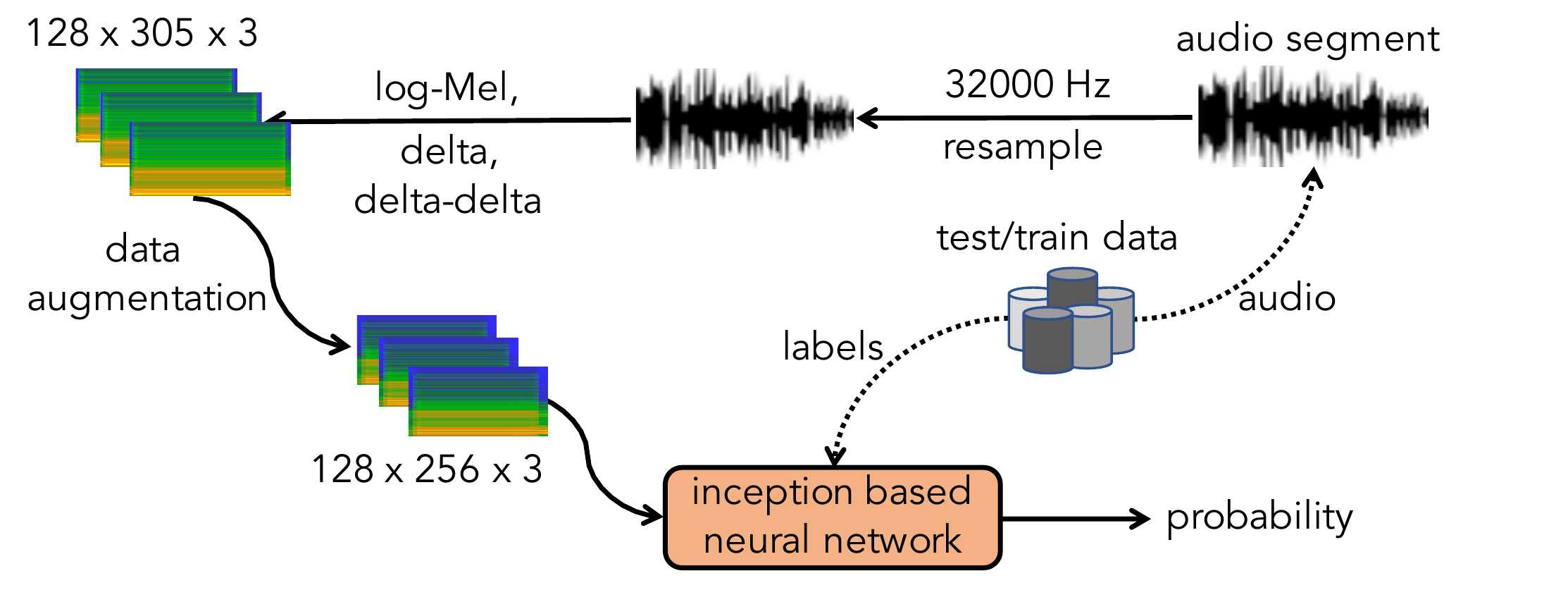}
	\caption{The high-level architecture of the proposed ASC baseline.}
    \label{fig:baseline_pic}
\end{figure}
\begin{table}[t]
	\caption{The inception based neural network used for the classification step in the ASC baseline system} 
	\centering
    \scalebox{0.8}{
	\begin{tabular}{|c |c| l |}
		\hline 
		\textbf{Main Blocks}         & \textbf{Sub blocks} & \textbf{Layers}   \\  
		\hline 
		\multirow{2}{*}{CNN-based} & Inception Block 01& Inc($Ch$=128)  - BN - ReLU - MP[2$\times$2] - Dr(0.1)    \\                  
        \multirow{2}{*}{backbone}  & Inception Block 02& Inc($Ch$=128) - BN - ReLU - MP[2$\times$2] - Dr(0.15)   \\                      
  		     	                   & Inception Block 03& Inc($Ch$=256) - BN - ReLU - MP[2$\times$2] - Dr(0.2)   \\                        
		                	       & Inception Block 04& Inc($Ch$=256) - BN - ReLU - GMP - Dr(0.25)     \\                                
     	\hline 
  		  MLP-based                & Dense Block 01 & FC($Ch$=1024) - BN - ReLU - Dr(0.25)   \\                            
  		  classification           & Dense Block 02 & FC($Ch=C$) - Softmax             \\                               		\hline 
	\end{tabular}  
	}  
	\label{table:baseline_network} 
\end{table} 
To evaluate whether to extend deep neural network architectures in depth with trunks of convolutional layers is effective for ASC task, we firstly propose an ASC baseline system which presents an inception based architecture and low footprint model.
Next, we construct benchmark neural networks of VGG16, VGG19, ResNet50V2, ResNet152V2, DenseNet169, DenseNet201, and Xception, which presents much deeper convolutinal layers compared to the ASC baseline.
The proposed low-complexity ASC baseline is evaluated and compared with the benchmark and high-complexity architectures on DCASE 2020 Task 1A Development dataset.

As Figure~\ref{fig:baseline_pic} shows, the proposed ASC baseline framework is separated into three main steps: Front-end spectrogram feature extraction, online data augmentations, and back-end inception based deep neural network for classification.

\subsection{Front-end spectrogram feature extraction}
\label{frontend}
The input audio recordings are firstly resampled to 32,000 Hz. 
Then, the resampled audio recordings are transformed into Mel spectrograms by Librosa toolbox~\cite{librosa_tool}. 
As we set the Fast Fourier Transform (FFT) number, Hann window size, the hop size, and the filter number to 4096, 2048, 1024, and 128 respectively, a two-dimensional Mel spectrogram of 128$\times$312 is generated from one 10-second audio recording.
We finally apply delta and delta-delta on each two-dimensional spectrogram, create three-dimensional spectrogram of 128$\times$305$\times$3 (i.e., The channel dimension is three which is created by concatenating the original Mel spectrogram, delta, and delta-delta).

\subsection{Online data augmentations}
\label{augmentation}

\begin{figure}[t]
    \centering
    \includegraphics[width =0.9\linewidth]{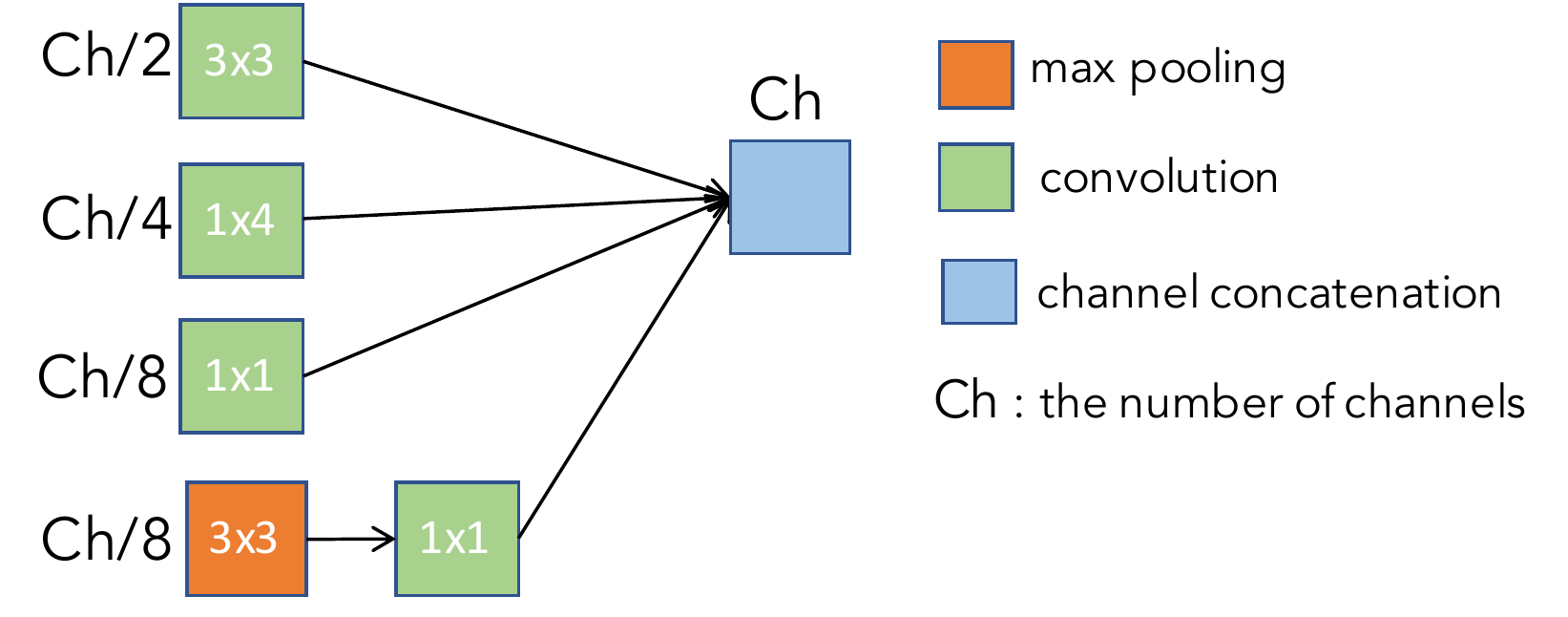}
	\caption{The architecture of the inception layer (Inc($Ch$ = The channel number) in the proposed ASC baseline.}
    \label{fig:baseline_inc_layer}
\end{figure}
%
In this paper, we apply three data augmentation methods: Random Cropping~\cite{aug_crop}, Specaugment~\cite{aug_spec}, and Mixup~\cite{mixup1, mixup2}, respectively. 
Firstly, the temporal dimension of Mel spectrograms of 128$\times$\textbf{305}$\times$3 is randomly cropped to 128$\times$\textbf{256}$\times$3 (Random Cropping). 
Next, ten random and continuous temporal and frequency bins of the cropped spectrograms are erased (Specaugment). 
Finally, the spectrograms are randomly mixed together using different coefficients from both Beta and Uniform distributions (Mixup). 
As all of three data augmentation methods are applied on each batch of spectrograms in the training process, we refer them to as the online data augmentations.
\begin{table*}[t]
    \caption{Compare our proposed ASC baseline to DCASE baseline, benchmark network architectures on the DCASE 2020 Task 1A Development sets across the different recording devices (Acc.\%)} 
        	\vspace{-0.2cm}
    \centering
    \scalebox{0.8}{

    \begin{tabular}{|c||c||c|| c|c||c|c|| c|c||  c|c||c|} 
    \hline 
	        \textbf{Performances} &\textbf{DCASE}  &\textbf{Proposed}   &\textbf{MobileV1}  &\textbf{MobileV2}  &\textbf{VGG16}   &\textbf{VGG19}  &\textbf{ResNet50V2} &\textbf{ResNet152V2} &\textbf{DenseNet121} &\textbf{DenseNet201} &\textbf{Xception} \\
	         &\textbf{Baseline} &\textbf{Baseline}  & & & & & &   & &  & \\  
     \hline 
     \hline        
        \textbf{A(\%)}                &70.6  &73.3  &74.2 &71.0 &68.3       &67.1    &74.1  &74.0  &74.1  &74.8  &75.2\\
        \textbf{B(\%)}                &60.6  &67.0  &60.1 &56.3 &54.5       &56.1    &57.9  &60.8  &63.1  &58.3  &62.0\\
        \textbf{C(\%)}                &62.6  &72.6  &63.7 &60.6 &61.5       &61.5    &63.7  &67.8  &63.7  &68.6  &68.4\\
        \textbf{S1(\%)}               &55.0  &64.2  &57.2 &52.5 &55.3       &49.8    &60.2  &52.5  &62.0  &57.2  &60.1\\
        \textbf{S2(\%)}               &53.3  &64.9  &51.4 &55.2 &54.4       &51.4    &54.1  &52.8  &58.9  &56.3  &54.7\\
        \textbf{S3(\%)}               &51.7  &67.9  &55.4 &52.5 &53.5       &52.0    &55.6  &57.0  &60.2  &59.6  &62.2\\
        \textbf{unseen-S4(\%)}        &48.2  &57.6  &43.8 &41.3 &43.8       &38.3    &45.6  &47.6  &51.7  &51.5  &50.4\\
        \textbf{unseen-S5(\%)}        &45.2  &60.0  &44.7 &46.1 &45.3       &44.4    &52.0  &44.3  &53.8  &48.7  &49.4\\
        \textbf{unseen-S6(\%)}        &39.6  &52.4  &32.6 &29.5 &40.1       &31.4    &31.7  &31.9  &40.8  &35.7  &35.2\\ 
        \hline   
        \hline       
        \textbf{Average(\%)}          &54.1  &64.6  &53.3 &51.6 &53.3       &50.8    &55.1 &54.0  &58.7  &56.7  &57.9\\         
        \hline
        \hline
        \textbf{Parameters(M)}        &5.0   &2.8   &4.3  &3.5  &15.3       &20.6    &25.7 &60.5  &8.1   &20.3  &23.0 \\   
        \textbf{Memory(MB)}           &19.2  &10.6  &16.4 &13.7 &58.2       &254.8   &98.0 &230.6 &30.9  &77.5  &87.6 \\  
    \hline 

    \end{tabular}
    }
    \label{table:baseline_res} 
\end{table*}
\subsection{Back-end inception based deep neural network}
\label{backend}

As Table~\ref{table:baseline_network} shows, the back-end classification is separated into two main parts: CNN-based backbone and Multilayer Perception (MLP) based classification.
In particular, the inception-based backbone comprises four Inception Blocks, each of which is performed by an inception layer, followed by batch normalization (BN)~\cite{batchnorm}, Rectified Linear Unit (ReLU)~\cite{relu}, drop out (Dr(drop ratio))~\cite{dropout}, and Max Pooling (MP) for the first three Inception Blocks or Global Max Pooling (GMP) for the final Inception Block 04.
The inception layer architecture (Inc($Ch$ = The channel number)) is shown in Figure~\ref{fig:baseline_inc_layer} which is a variant of the naive version of inception layer introduced in~\cite{inception_ref}.
In particular, we use kernel [1$\times$4] instead of [5$\times$5] as usual to enforce the network focus on minor variation across the frequency dimension of audio spectrum. 
Additionally, we add a convolutional layer with kernel size of [1$\times$1] after the max pooling MP([3$\times$3]) layer.

Regarding the MLP-based classification as shown the lower part in Table~\ref{table:baseline_network}, it performs two dense blocks (Dense Block 01 and Dense Block 02). 
While the fully connected layers (FC($Ch$ = The channel number)) in the first Dense Block 01 is followed by Rectified Linear Unit (ReLU) and drop out (Dr(drop ratio), the fully connected layer (FC($Ch$ = $C$)) in the second Dense Block 02 uses Softmax layer (i.e., $C$ is set to match the number of categories classified in a target dataset.)

\subsection{Construct benchmark and high-complexity neural networks for back-end classification}
\label{baseline_benchmark}
To evaluate whether benchmark and high-complexity deep neural network architectures are effective for ASC task, we replace the proposed CNN-based backbone in the ASC baseline model by the network architectures of MobileNetV1, MobileNetV2, VGG16, VGG19, ResNet50V2, ResNet152V2, DenseNet169, DenseNet201, and Xception which are available from Keras Application API~\cite{keras_app}.
In other words, while the proposed MLP-based classification is remained, we evaluated different backbone network architectures.
Notably, steps of the front-end spectrogram feature extraction and the online data augmentations used for the ASC baseline are retained during evaluating these benchmark network architectures. 

\subsection{Dataset and settings for evaluating the proposed ASC baseline and the benchmark neural networks}
\label{baseline_dataset}
To evaluate the proposed ASC baseline and the benchmark neural networks, we use DCASE 2020 Task 1A Development dataset mentioned in Section~\ref{data_set}.
We obey the challenge and follow the recommended setting as mentioned in Section~\ref{data_set}. 
Regarding the evaluation metric, we use accuracy (Acc.\%), which is the most popular and main metric in all ASC challenges~\cite{dcase_community}.

As using the Mixup data augmentation method, labels are not one-hot encoding format. 
Therefore, we use Kullback–Leibler divergence (KL) loss~\cite{kl_loss} shown in Eq. (\ref{eq:kl_loss}) below.
\begin{align}
   \label{eq:kl_loss}
   Loss_{KL}(\Theta) = \sum_{n=1}^{N}\mathbf{y}_{n}\log \left\{ \frac{\mathbf{y}_{n}}{\mathbf{\hat{y}}_{n}} \right\}  +  \frac{\lambda}{2}||\Theta||_{2}^{2}
\end{align}
where  \(\Theta\) are trainable parameters, constant \(\lambda\) is set initially to $0.0001$, $N$ is batch size set to 64, $\mathbf{y_{i}}$ and $\mathbf{\hat{y}_{i}}$  denote expected and predicted results.

Both the proposed ASC baseline and the benchmark neural networks are implemented with Tensorflow framework, using Adam~\cite{Adam} for optimization. 
The training and evaluating processes are conducted on GPU Titan RTX 24GB
The training process is stop after 40 epoches.
While the first 30 epoches uses the learning rate of 0.001 and all data augmentation methods mentioned in Section~\ref{augmentation}, the remaining epoches uses the lower learning rate of 0.00001 with only Random Cropping data augmentation method. 
The final result, which is reported in this paper, is an average of accuracy after 10 times of running experiments.
\begin{figure*}[t]
    \centering
    \includegraphics[width =1.0\linewidth]{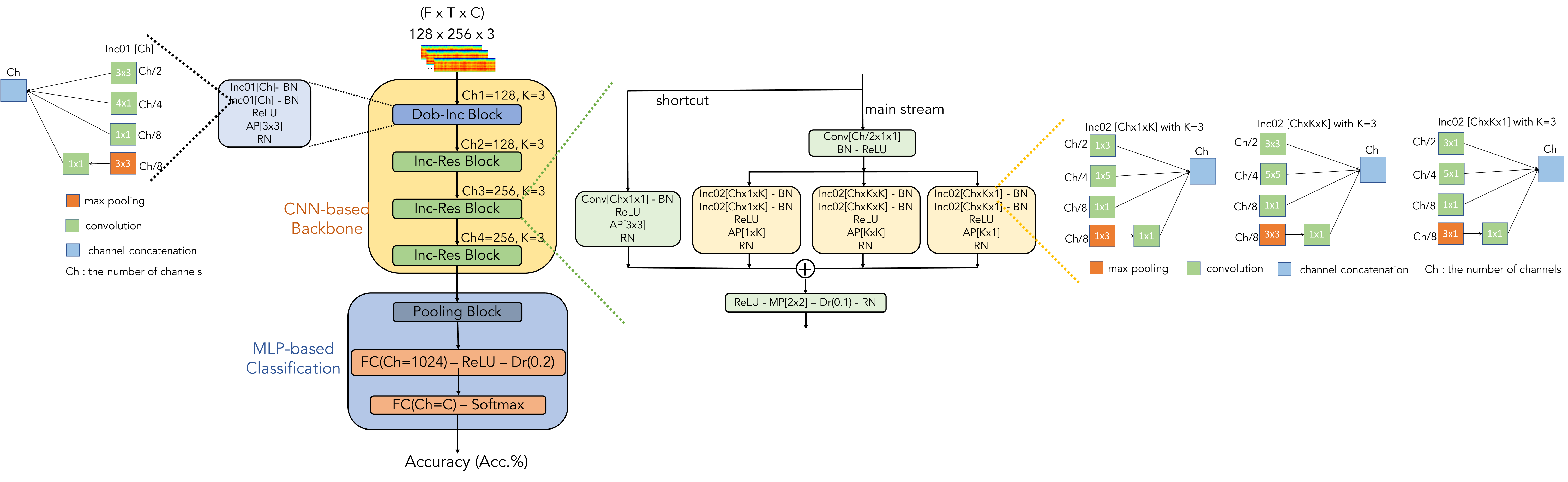}
	\caption{The proposed novel residual-inception deep neural network architecture.}
   	\vspace{-0.4cm}
    \label{fig:novel_pic}
\end{figure*}
\subsection{Performance comparison among DCASE baseline, the proposed ASC baseline, and the benchmark network architectures} 
\label{baseline_per}
As experimental results are shown in Table~\ref{table:baseline_res}, our proposed ASC baseline system outperforms DCASE baseline and all the benchmark network architectures.
Significantly, our proposed ASC baseline helps to improve the DCASE baseline on all seen and unseen recording devices.
The results of the proposed ASC baseline also indicate that performance on unseen devices (S4, S5, and S6) are lower than seen devices (A, B, C, S1, S2, and S3) an average of 10\% and the performance on real recording devices (A, B, C) are better than synthesized devices (S1 to S6).

Regarding the number of trainable parameters used in evaluating models, we can see that deeper neural networks (MobileNetV1, VGG19, ResNet152V2 or DenseNet201) present low performance than the lower complexity networks (MobileNetV2, VGG16, ResNet50V2, or DenseNet121) from the same architecture groups. 
Meanwhile, our proposed ASC baseline presents a low footprint of 2.8 M trainable parameters, but achieve the best performance compared to the others.
We also see that only DCASE baseline, our ASC baseline, and MobinetV1/V2 network architectures present lower than 5 M trainable parameters while the others show high-complexity models.
As one trainable parameter is presented by 32 bits using floating point format in this paper, the proposed ASC baseline occupies 10.6 MB memory on devices.

Overall, these experimental results indicate that the shallow inception based neural network used for the ASC baseline is more effective than deeper architectures for ASC task with the issue of mismatched recording devices.
Although the ASC baseline presents the lowest footprint model with 10.6 MB memory occupation which is compatible to a wide range of mobiles or edge devices, this network architecture is still considered too large regarding small and limited-memory devices such as STM32L496@80MHz or Arduino Nano 33@64MHz.
Additionally, the performance of the proposed ASC baseline (64.6\%) is not competitive to the state-of-the-art systems. 
Therefore, these below sections will show how we improve the ASC baseline performance, but still satisfy the requirement of low-complexity model.

\section{Propose a novel residual-inception neural network and an ensemble of multiple spectrograms to improve the ASC baseline}
\label{novel_section}

As the performance comparison between the proposed ASC baseline and the benchmark neural networks are shown in Section~\ref{baseline_section}, it indicates that the inception-based architecture shows effective for ASC task. 
To further improve the ASC performance, we therefore explore the inception-based architecture, then propose a novel residual-inception (NRI) neural network as shown in Figure~\ref{fig:novel_pic}.

\subsection{Propose a novel residual-inception network architecture}
\label{novel_arc}

As Figure~\ref{fig:novel_pic} shows, the proposed residual-inception network architecture also comprises two main parts: CNN-based deep neural network backbone and multiplayer perception (MLP) based classification.
In particular, there are four blocks in the proposed CNN-based backbone: one Dob-Inc Block and three Inc-Res Blocks.
These four blocks are described at the upper part of Figure~\ref{fig:novel_pic}.
While Dob-Inc Block makes uses of inception-based architecture, both inception-based and residual architectures are leveraged in three Inc-Res Blocks.

Regarding the Dob-Inc Block as shown in the left part of Figure~\ref{fig:novel_pic}, it reuses the sub-block architecture of Inception Block from the ASC baseline, but using two Inc01(Ch) layers each of which is accompanied with batch normalization (BN). 
The number of channel (Ch) used at these inception layers is set to 128.
Three Inc-Res Blocks as shown in the right part of Figure~\ref{fig:novel_pic} present the same network architecture, and the channel numbers are set to 128, 256, and 256, respectively. 
Each Inc-Res Block presents two data streams.
As the first stream is shown on the left, referred to as the shortcut branch, the feature map input goes through (Conv[Ch$\times$1$\times$1]), BN, ReLU, AP[3$\times$3], Residual Normalization (RN($\lambda=0.4$)) inspired from~\cite{kim2021qti}.
Meanwhile, in the main stream as shown on the right, the feature map input firstly goes though (Conv[Ch/2$\times$1$\times$1]), BN, and ReLU, then is passed into three sub branches.
In each sub branch in the main stream, different kernel sizes, defined by $K$ as shown in the right of Figure~\ref{fig:novel_pic}, are applied to explore local feature of the feature map input.
By using different kernel sizes of [K$\times$1], [K$\times$K], and [1$\times$K] and applying AP layers with the same kernels [K$\times$K], the network is enforced to learn distribution of spectrum in certain frequency bands effectively.
This strengthens the network to deal with the issue of mismatched recording devices which make the different distribution of energy across the frequency dimensions.
Finally, the shortcut stream and three sub branches in the main stream are accumulated before going through ReLU, MP[2$\times$2], Dr(0.1), and RN($\lambda=0.4$) in the order.

The MLP-based classification as shown in the lower part of Figure~\ref{fig:novel_pic} performs a Pooling Block and two fully connected layer blocks.
At the Pooling Block, thee types of feature are extracted: (1) global average pooling across the channel dimension (i.e., this is exactly the global average pooling layer (GAP) used in the proposed ASC baseline), (2) global max pooling across temporal dimension, and (3) global average pooling across frequency dimension.
We then concatenate these features before feeding into fully connected blocks.
While the first fully connected layer (FC(Ch=1024)) is followed by ReLU and Dr(0.2), the second fully connected layer combines with Softmax layer for classifying into $C$ scene categories.

\subsection{Further improve ASC performance by an ensemble of multiple spectrogram inputs}
\label{novel_ens}
As using ensemble is a rule of thumb to improve the ASC performance and shows effective to deal with the issue of mismatched recording devices~\cite{lam_ijcnn, truc_dca_18, mul_spec_2020, lam_dca_18, lam_dca_16_int, huy_mul, pham_ict}, we therefore apply an ensemble of multiple spectrogram inputs in this paper.
In particular, we use three spectrograms: log-Mel~\cite{librosa_tool}, Gammatone (Gam)~\cite{auditory2009_tool}, and Constant Q Transform (CQT)~\cite{librosa_tool}. 
To make sure the spectrograms present the same size, we reuse the setting parameters of FFT number, Hann window size, the hop size, and the filter number as mentioned in Section~\ref{frontend} and apply for three spectrograms.
As using multiple spectrogram inputs, each of spectrogram is trained with one back-end deep learning model. 
Then, predicted probabilities obtained from individual models will be fused to achieve the best performance.
In this paper, we propose to use late fusion of probabilities, referred to as PROD fusion.
Let consider predicted probabilities of each model as  \(\mathbf{\bar{p_{s}}}= (\bar{p}_{s1}, \bar{p}_{s2}, ..., \bar{p}_{sC})\), where $C$ is the category number and the \(s^{th}\) out of \(S\) networks evaluated. 
Next, the predicted probabilities after PROD fusion \(\mathbf{p_{prod}} = (\bar{p}_{1}, \bar{p}_{2}, ..., \bar{p}_{C}) \) is obtained by:
\begin{equation}
\label{eq:mix_up_x1}
\bar{p_{c}} = \frac{1}{S} \prod_{s=1}^{S} \bar{p}_{sc} ~~~  for  ~~ 1 \leq c \leq C 
\end{equation}
Finally, the predicted label  \(\hat{y}\) is determined by 
\begin{equation}
    \label{eq:label_determine}
    \hat{y} = arg max (\bar{p}_{1}, \bar{p}_{2}, ...,\bar{p}_{C} )
\end{equation}

\subsection{Dataset and settings for evaluating the novel residual-inception neural network}
\label{novel_dataset}
To evaluate the novel residual-inception network architecture and how an ensemble of multi-spectrogram inputs can help to further improve performance, we use DCASE 2020 Task 1A Development dataset.
All settings and implementation are reused from Section~\ref{baseline_dataset}.
\begin{table}[t]
   \caption{Performance comparison among: DCASE baseline, our ASC baseline, the novel residual-inception network (NRI) with individual spectrograms of Mel (Mel-NRI), Gam (Gam-NRI), or CQT (CQT-NRI), and ensemble of multiple spectrograms (SPECs-NRI) on DCASE 2020 Task 1A Development dataset} 
        	\vspace{-0.2cm}
    \centering
    \scalebox{0.85}{

    \begin{tabular}{|c|  c|c| c|c|c|  c| } 
        \hline 
	     &\textbf{DCASE}    &\textbf{Proposed} &\textbf{Mel-}  &\textbf{Gam-}  &\textbf{CQT-} &\textbf{SPECs-} \\
	     &\textbf{baseline} &\textbf{baseline} &\textbf{NRI}  &\textbf{NRI}    &\textbf{NRI}  &\textbf{NRI} \\

        \hline 
        \hline                                                  
       
        \textbf{A(\%)}            &70.6 &73.3 &77.3 &77.3  &61.5            &80.6  \\
        \textbf{B(\%)}            &60.6 &67.0 &70.5 &67.8  &62.3            &78.7  \\
        \textbf{C(\%)}            &62.6 &72.6 &75.7 &71.1  &57.1            &73.9  \\
        \textbf{S1(\%)}           &55.0 &64.2 &69.7 &68.2  &62.1            &74.6  \\
        \textbf{S2(\%)}           &53.3 &64.9 &70.6 &59.4  &63.6            &74.2  \\
        \textbf{S3(\%)}           &51.7 &67.9 &71.8 &70.9  &61.2            &76.4  \\
        \textbf{unseen-S4(\%)}    &48.2 &57.6 &61.5 &61.2  &60.9            &67.3  \\
        \textbf{unseen-S5(\%)}    &45.2 &60.0 &66.1 &63.3  &60.3            &71.8  \\
        \textbf{unseen-S6(\%)}    &39.6 &52.4 &58.8 &52.7  &58.1            &65.2  \\
        \hline    
        \hline                                                                                                                                                                                   
        \textbf{Average(\%)}      &54.1 &64.6 &69.1 &65.8  &60.8            &73.6  \\  
        \hline   
               \hline                                                  
                                              
        \textbf{Parameters (M)}    &5.0    &2.8   &4.3    &4.3   &4.3       &12.9  \\  
        \textbf{Memory (MB)}      &19.2   &10.6  &16.6   &16.6  &16.6       &49.8  \\  
       \hline 
    \end{tabular}
    }
    \label{table:novel_res} 
\end{table}

\subsection{Performance comparison among DCASE baseline, the proposed ASC baseline, the novel residual-inception network architecture with individual spectrograms, and the ensemble of multiple spectrograms}
\label{novel_per}
As experimental results are shown in Table~\ref{table:novel_res}, we can see that the novel residual-inception (NRI) network trained with Mel spectrogram (Mel-NRI) helps to further improve the proposed ASC baseline in Section~\ref{baseline_section} by 4.5\% and significantly outperform DCASE baseline with an improvement of 15.0\%.

Compare among spectrograms, the novel residual-inception networks trained with Mel spectrogram (Mel-NRI) and Gam spectrogram (Gam-NRI) are competitive, presenting classification accuracy of 69.1\% and 65.8\%, respectively.
Ensemble of these two model achieves an accuracy of 69.9\%, slightly improve Mel-NRI by 0.8\%.
Although the novel residual-inception networks trained on CQT spectrogram (CQT-NRI) presents a low performance of 60.8\%, ensemble of three spectrograms (SPECs-NRI) achieves an accuracy of 73.6\%, further improve the ensemble of Mel-NRI and Gam-NRI by 3.7\%.

Regarding performance on different recording devices, SPECs-NRI ensemble model significantly improves the performance on all recording devices, further improve DCASE baseline and the ASC baseline an average of 23\% and 10\%, respectively.
The gap performance between real recording devices (A, B, C) and synthetic devices (from S1 to S6) as well as between unseen devices (S4, S5, S6) and seen devices (A, B, C, S1, S2, S3) are also reduced by using ensemble of three spectrograms (SPECs-NRI). 
Again, we have proven that ensemble of multi-spectrogram inputs is effective for ASC system to deal with the issue of mismatched recording devices.

\section{Analyze the trade off between ASC model complexity and performance}
\label{tradeoff_section}

\subsection{Propose techniques to reduce the model complexity}
\label{tradeoff_comp}
As Table~\ref{table:novel_res} shows, the complicated architecture of the novel residual-inception (NRI) network and applying an ensemble of multiple spectrograms cause increasing the number of trainable parameters to 12.9 M, occupying 49.8 MB memory on devices.
As we aim to achieve low-complexity ASC systems which are suitable for various edge devices, two constrains of the maximum memory occupied by proposed models are set: (1) 20 MB for mobiles or large-memory devices basing on surveys in~\cite{model_size_sur_01, model_size_sur_02}, and (2) 128 KB for limited-memory devices such as STM32L496@80MHz or Arduino Nano33@64MHz (i.e., These limited-memory devices normally does not support operation system (OS)). The second constrain of 128 KB memory is also the requirement set to challenges of DCASE 2021 Task 1A and DCASE 2022 Task 1.
The models with the memory constrains of 20 MB and 128 KB are referred to as medium-size model (MM) and small-size model (SM) respectively in this paper.
Meanwhile, models with unlimited memory occupation is referred to as large-size model (LM).  

\begin{figure}[t]
    \centering
    \includegraphics[width =1.0\linewidth]{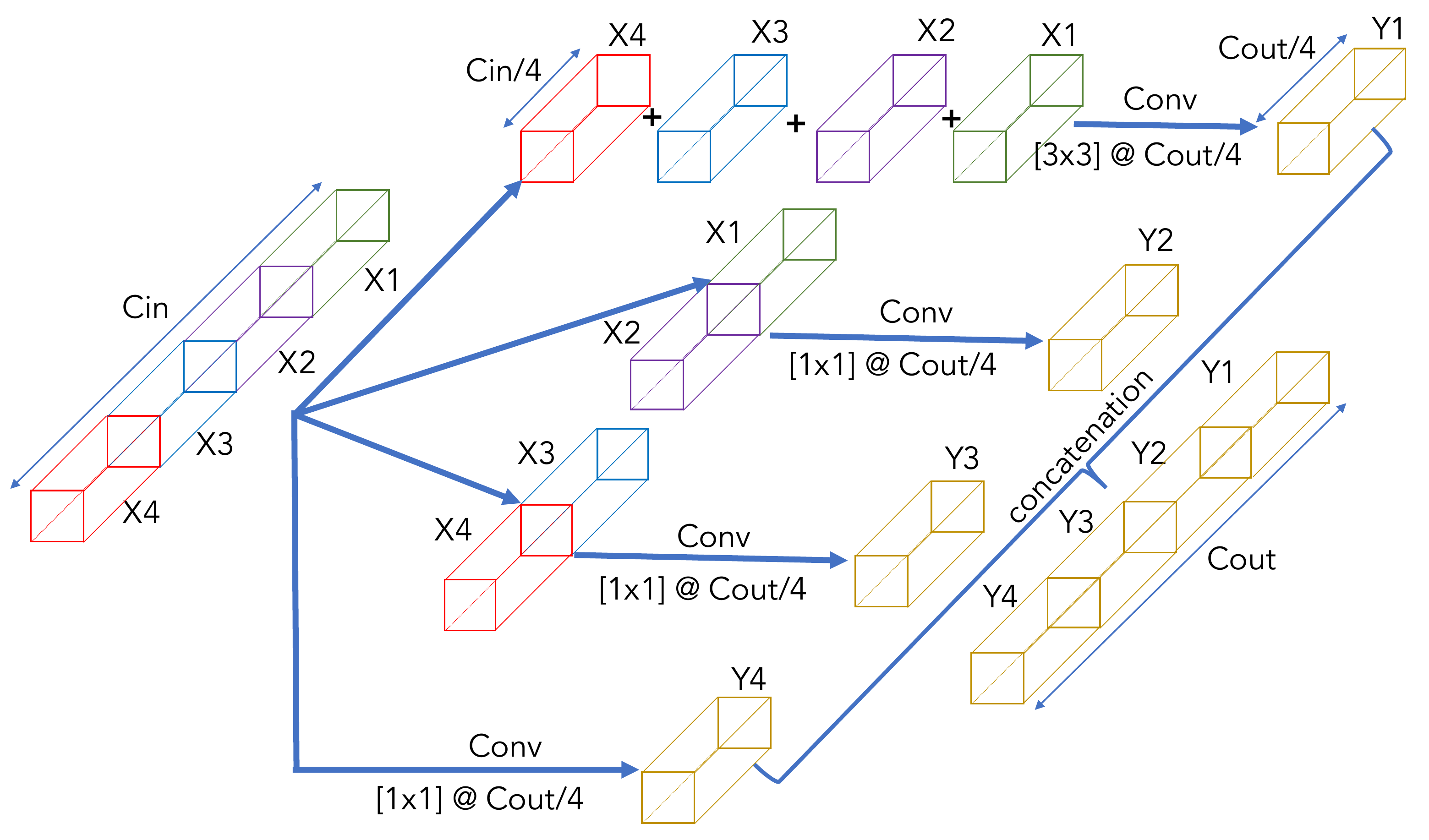}
	\caption{Channel deconvolution (CD) for reducing trainable parameters}
    \label{fig:novel_dec}
\end{figure}

To achieve these low-complexity models (MM and SM), we apply three techniques of model compression: Channel reduction (CR), channel deconvolution (CD), and quantization (Qu).
In particular, we firstly reduce the channel number in inception layers at each sub block of NRI network to 128, 64, and 32 respectively.
To further reduce the model complexity, the channel deconvolution technique is applied on each convolutional layer with a kernel size of [K$\times$K], which is inspired from~\cite{dec_channel_01, dec_channel_02}.
Let consider $Cin$ and $Cout$ are the number of input channel and output channel used at a convolutional layer with kernel size of [K$\times$K] as shown in Figure~\ref{fig:novel_dec}.
We then separate the input tensor basing on the length of channel dimension ($Cin$) into 4 sub-tensors: 0 to $Cin/4$, $Cin/4$ to $Cin/2$, $Cin/2$ to $3.Cin/4$, and $3.Cin/4$ to $Cin$, referred to as $X1$, $X2$, $X3$, and $X4$, respectively.
Then, convolutional layers with different kernel sizes as shown in Figure~\ref{fig:novel_dec} are applied to learn these sub-tensors before concatenating.
By using the channel deconvolution (CD), the trainable parameters used for a convolutional layer with kernel [K$\times$K] is reduced to nearly 1/8.5 of the original number of trainable parameters.

By using both channel reduction (CR) and channel deconvolution (CD), we have Table~\ref{table:novel_red} which summaries four variants of SPECs-NRI, referred to as SPECs-NRI-RD128, SPECs-NRI-RD64, SPECs-RNI-RD32, and SPECs-RNI-120KB with the low complexity of 2.62 M, 0.86 M, 0.36 M, and 0.12 M of trainable parameters, respectively.
We can see that these four variants of SPECs-NRI are considered as medium-size model (MM) which meet the requirement of maximum 20 MB of memory occupation on devices.
To meet the requirement of maximum 128 KB of memory, we apply the quantization technique to only SPECs-RNI-120KB (i.e., the quantization technique helps to convert 32-bit floating point to 8-bit integer, which reduce the memory occupation to 1/4 of the original volume), further reduce the occupied memory from 0.48 MB to 120 KB. 

\subsection{Dataset and settings for evaluating the techniques of model compression}
\label{tradeoff_dataset}
To evaluate the techniques of model compression applied on the proposed residual-inception neural network, we again uses DCASE 2020 Task 1 Development dataset. 
All settings and implementation are reused from Section~\ref{baseline_dataset}.

\subsection{Performance of the novel residual-inception network architecture with and without using decompression techniques}
\label{tradeoff_per}

As the experimental results are shown in Table~\ref{table:tradeoff_res}, we can see that applying model compression techniques CR and CD on SPECs-NIR to slightly decreases the classification accuracy.
In particular, variants of SPECs-NRI-RD128, SPECs-NRI-RD64, SPECs-NRI-RD32 present the accuracy of 72.9\%, 72.0\%, 71.3\%, respectively compared with 73.6\% of SPECs-NRI model.
However, these techniques helps to reduce the model complexity significantly, presenting 9.9 MB, 3.3 MB, and 1.4 MB memory occupation for SPECs-NRI-RD128, SPECs-NRI-RD64, SPECs-NRI-RD32, respectively.

Further apply the quantization technique on SPECs-RNI-120KB, we achieve a very low-complexity model of 120 KB, but still perform an accuracy of 71.0\%.
\begin{table}[t]
    \caption{Channel numbers and model complexities after applying channel reduction (CR) and channel deconvolution (CD)} 
        	\vspace{-0.2cm}
    \centering
    \scalebox{0.7}{

    \begin{tabular}{|l|  c|c| c |c | c|} 
        \hline 
	      \textbf{Sub blocks}      & \textbf{SPECs-NRI}   & \textbf{SPECs-NRI} &\textbf{SPECs-NRI}  &\textbf{SPECs-NRI}  &\textbf{SPECs-NRI}\\
	                               &                      & \textbf{-RD128}    &\textbf{-RD64}      &\textbf{-RD32}      &\textbf{-120KB}\\

        \hline          
        \hline        
        \textbf{Dob-Inc}                 &$Ch$=128  &$Ch$=128  &$Ch$=64   &$Ch$=32    &$Ch$=16 \\
        \textbf{Inc-Res 01}              &$Ch$=128  &$Ch$=128  &$Ch$=64   &$Ch$=32    &$Ch$=32\\
        \textbf{Inc-Res 02}              &$Ch$=256  &$Ch$=128  &$Ch$=64   &$Ch$=32    &$Ch$=32\\
        \textbf{Inc-Res 03}              &$Ch$=256  &$Ch$=128  &$Ch$=64   &$Ch$=32    &$Ch$=32\\
        \textbf{Fully connected 01}      &$Ch$=1024 &$Ch$=1024 &$Ch$=1024 &$Ch$=1024  &-\\
        \textbf{Fully connected 02}      &$Ch$=10   &$Ch$=10   &$Ch$=10   &$Ch$=10    &$Ch$=10\\
       \hline                                       
       \hline                                                                                           
       \textbf{Parameters (M)}            &12.9   &2.62  &0.86  &0.36  &0.12 \\       
       \textbf{Memory (MB)}               &49.8   &9.9   &3.3   &1.4   &0.48 \\  
       \hline                                                                                           

    \end{tabular}
    }
    \label{table:novel_red} 
\end{table}
\begin{table}[t]
    \caption{Performance comparison among SPECs-NRI and four variants of SPECs-NRI-RD128, SPECs-NRI-RD64, SPECs-NRI-RD32, and SPECs-NRI-120KB w/ quantization on DCASE 2020 Task 1A Development dataset} 
        	\vspace{-0.2cm}
    \centering
    \scalebox{0.7}{

    \begin{tabular}{|c|c|  c|c|c|c|} 
        \hline 
	         &\textbf{SPECs-NRI} &\textbf{SPECs-NRI}  &\textbf{SPECs-NRI} &\textbf{SPECs-NRI} &\textbf{SPECs-NRI}\\
   	         &                   &\textbf{-RD128}     &\textbf{-RD64}     &\textbf{-RD32}     &\textbf{-120KB}\\
        \hline 
               \hline 
        \textbf{A(\%)}             &80.6 &77.9 &76.4 &75.5 &75.7\\
        \textbf{B(\%)}             &78.7 &75.4 &74.5 &72.3 &72.0\\
        \textbf{C(\%)}             &73.9 &78.1 &73.9 &74.8 &77.2\\
        \textbf{S1(\%)}            &74.6 &71.5 &73.3 &73.0 &69.9\\
        \textbf{S2(\%)}            &74.2 &77.3 &74.5 &70.9 &69.9\\
        \textbf{S3(\%)}            &76.4 &76.7 &73.6 &75.1 &74.8\\
        \textbf{unseen-S4(\%)}     &67.3 &67.3 &68.5 &70.9 &69.6\\
        \textbf{unseen-S5(\%)}     &71.8 &69.7 &71.8 &70.6 &71.2\\
        \textbf{unseen-S6(\%)}     &65.2 &62.4 &62.1 &58.8 &61.5\\
        \hline   
        \hline                                                                       
        \textbf{Aver.(\%)}         &73.6 &72.9 &72.0 &71.3 &71.0\\  
       \hline                                       
       \hline                                                                                           
        \textbf{Parameters (M)}            &12.9   &2.62  &0.86  &0.36  &0.12 \\       
       \textbf{Memory (MB)}               &49.8   &9.9   &3.3   &1.4   &0.12 \\   
       \hline 
    \end{tabular}                  
    }
    \label{table:tradeoff_res} 
\end{table}
\section{Explore sound event information to further improve the ASC performance}
\label{event_section}

\subsection{Proposed method to use sound event information to improve ASC performance}
\label{event_method}

To further improve ASC performance by leveraging sound event information, we firstly define the task of sound event detection (SED) as the up-stream task where sound events in a sound recording are detected.
The available models used for SED task is called as the up-stream pre-trained models.
We then leverage the pre-trained models, feed spectrograms of sound scene recordings into these models to extract certain feature map, referred to as the audio-event-based embeddings.
These embeddings are finally classified by a MLP based network into target sound scene classes.
In other words, classifying the audio-event-based embeddings using MLP based network is considered as the down-stream ASC task.
To the best our knowledge, there are thee papers proposed various up-stream pre-trained models which was trained on the AudioSet, the largest Audio dataset of sound events.
The first paper published by Google introduced Trill model~\cite{trill_model} and Frill model~\cite{frill_model} which reused the MobileNetV3 and ResNet50 architectures, respectively.
These two pre-trained models present the trainable parameters of 98.1 M and 38.5 M, respectively.
The second paper introduced a VGGish network architecture, referred to as openL3 model~\cite{openl3_model, openl3_model_02}, which presents 5.3 M trainable parameters.
Meanwhile, the third paper~\cite{pann_paper} presented a wide range of up-stream pre-trained networks using VGGish, ResNet, MobileNet, DaiNet, LeeNet, Res1dNet, and Wavegram based architectures.
As we aim to achieve a low complexity model less than 5 M of trainable parameters or approximately occupying 20 MB memory on devices and mobiles in this paper, we therefore reuse the pre-trained MobinetV2 network from~\cite{pann_paper} which presents the smallest footprint of 4.1 M trainable parameters (occupying 16.0 MB memory on devices). 
Notably, as all available up-stream pre-trained models recently mentioned are larger than 15 MB, we do not aim to achieve a low complexity model with 128 KB memory occupation with leveraging sound event information in this section.

Given the up-stream pre-trained MobileNetV2 model in~\cite{pann_paper}, we feed Mel spectrograms of sound scene recordings into this model to extract sound-event-based embeddings. 
The extracted embeddings are the feature map at the global pooling layer of the up-stream pre-trained MobileNetV2 model.
We then use the MLP-based classifier as shown in Table~\ref{table:baseline_network} to classify these sound-event-based embeddings into $C$ target sound scene categories.
This down-stream ASC task is referred to as DS-ASC-MobV2.
The predicted probabilities from the down-stream ASC task, DS-ASC-MobV2, is finally fused with the probabilities obtained from the novel residual-inception based network of SPECs-NRI-RD64 (i.e., the PROD fusion method as mentioned in Section~\ref{novel_ens} is used to fuse the probability results).
Notably, as the pre-trained MobileNetV2 and SPECs-NRI-RD64 networks present 4.1 M and 0.86 M of trainable parameters, the ensemble of these models presents 4.96 M of trainable parameter which satisfies our target of low complexity ASC model less than 5 M of trainable parameters or occupying 20 MB memory in this paper.
\begin{table}[t]
    \caption{Performance comparison among MEL-NIR, SPECs-NRI-RD64, Down-stream ASC task (DS-ASC-MobV2), and ensemble of SPECs-NRI-RD64 and DS-ASC-MobV2 on DCASE 2020 Task 1A Development dataset} 
        	\vspace{-0.2cm}
    \centering
    \scalebox{0.7}{

    \begin{tabular}{|c|  c|c| c|c|} 
        \hline 
                                    &\textbf{SPECs-NRI} &\textbf{SPECs-NRI-RD64} &\textbf{DS-ASC-MobV2}   &\textbf{DS-ASC-MobV2,} \\
                        	        &                 &                        &                        &\textbf{SPECs-NRI-RD64}         \\

        \hline 
       
        \textbf{A(\%)}           &80.6               &76.4                    &65.8     &78.8\\
    	\textbf{B(\%)}           &78.7               &74.5                    &58.7     &74.5\\
        \textbf{C(\%)}           &73.9               &73.9                    &67.5     &79.6\\
        \textbf{S1(\%)}          &74.6               &73.3                    &54.9     &74.5\\
        \textbf{S2(\%)}          &74.2               &74.5                    &52.7     &76.9\\
        \textbf{S3(\%)}          &76.4               &73.6                    &57.0     &77.0\\
        \textbf{unseen-S4(\%)}   &67.3               &68.5                    &56.1     &68.8\\
        \textbf{unseen-S5(\%)}   &71.8               &71.8                    &58.2     &70.3\\
        \textbf{unseen-S6(\%)}   &65.2               &62.1                    &59.1     &64.8\\
     \hline                                               
     \hline                                                                                                                                                   
        \textbf{Aver.(\%)}       &73.6               &72.0                    &58.9     &73.9\\  
    \hline                                                       
     \hline                                                                                                                         
        \textbf{Parameters(M)}   &12.9               &0.86                    &4.1      &4.96  \\ 
        \textbf{Memory (MB)}     &49.8               &3.3                     &16.0      &19.4 \\  
       \hline                              
    \end{tabular}                  
    }
    \label{table:event_res} 
\end{table}

\subsection{Dataset and experimental settings to evaluate the role of sound events}
\label{event_dataset}

To evaluate the role of sound event information to improve ASC performance, we continue using DCASE 2020 Task 1A Development dataset.
All settings and implementation are reused from Section~\ref{baseline_dataset}.

\subsection{Performance of ASC models with using sound event information}
\label{event_per}
As experimental results are shown in Table~\ref{table:event_res}, the down-stream model of DS-ASC-MobV2 achieves an overall accuracy of 58.9\%.
The performance is nearly equal to CQT-NRI and significant lower than Mel-NRI and Gam-NIR.
It indicates that direct training on spectrogram input is better than the approach of using up-stream pre-trained models with the large-scale Audio dataset of sound event.

When we combine DS-ASC-MobV2 with SPECs-NRI-RD64, we can achieve a low complexity model with 4.96 M of trainable parameters (19.4 MB memory occupation).
The combination of DS-ASC-MobV2 and SPECs-NRI-RD64, medium-size model, outperforms the large-size model of SPECs-NRI (73.9\% compared to 73.1\%),  but present a lower model complexity (19.4 MB compared to 49.8 MB).
 
\begin{table}[t]
    \caption{527 sound events and alarming level definition from AudioSet dataset} 
        	\vspace{-0.2cm}
    \centering
    \scalebox{0.8}{

    \begin{tabular}{|c|l| } 
        \hline 
      \textbf{Levels}    &\textbf{Sound events} \\

        \hline 
        \textbf{Red Level}            &`Explosion', `Gunshot, gunfire', `Machine gun', \\ 
                                    &`Fusillade', `Artillery fire', `Cap gun', `Eruption', \\
                                    & `Fire', `Fireworks', `Firecracker' \\
                                            \hline 
       
        \textbf{Yellow Level}     &`Wail,moan', `Shout', `Bellow', `Whoop', `Yell', `Children  \\ 
        \textbf{(individual person)}                  &shouting', `Screaming', `Crying, sobbing', `Baby cry, infant cry'\\
                                        \hline 

        \textbf{Yellow Level}     &`Cheering', `Crowd', `Run', `Applause', `Hubbub, \\
        \textbf{(a crowd)}   & speech noise, speech babble', `Battle cry' \\
                                \hline
                                
        \textbf{Yellow Level}     &`Thunderstorm', `Thunder' \\
        \textbf{(from nature)} & \\         
        \hline
        
        \textbf{Yellow Level}     & `Basketball bounce', `Crackle', `Machanisms', \\
        \textbf{(thing and }      & `Detal drill', `Buzzer', `Hammer', `Jackhammer', \\
        \textbf{machine sounds)}  & `Power tool', `Drill', `Burst, pop', `Crack', `Skidding', \\
                                & `Toot', `Race car, auto racing', `Tire squeal', `Air brake',\\
                                & `Traffic noise, roadway noise', `Engine knocking', \\
                                & `Engine knocking', `Engine starting', `Breaking', `Bouncing', \\
                                & `Scratch', `Thump, thud', `Bang', `Slam', `Knock', `Tap' \\
                                \hline
         \textbf{Green Level}     & Other events \\
        
        \hline 

    \end{tabular}                  
    }
    \label{table:sound_event_level} 
\end{table}

\section{Propose a visualization method for well presenting a sound scene context}
\label{visual_section}

\subsection{Motivation for the developing a visualization method to present a sound scene context}
\label{visual_motivation}

The motivation to develop a visualization method to well present a sound scene context is driven from two main reasons. 
Firstly, as the literature review of ASC task in Section~\ref{intro} presents, current ASC tasks are specific, defined on certain datasets, and considered as an individual task.
Given an ASC model successfully trained on a dataset, we can only acknowledge that how an input audio recording is close to a certain scene context defined in the dataset basing on the predicted probabilities.
Therefore, when the ASC result is used as the input for other tasks in a complex system, it is hard to give a decision if the predicted probabilities are not significantly different.
Additionally, currently proposed ASC models have presented limited performances (i.e., the best models proposed for ASC tasks in DCASE challenges cannot achieve more than 90\% classification accuracy) as various challenges of mismatch recording devices, low-complexity model, or very similar sound scene contexts (e.g., pedestrian street and traffic street in DCASE challenges or firework event and riot context in Crowded Scenes dataset).    
As a result, applying ASC task as a main component or as a sub function in real-life applications is limited or show ineffective if only predicted probabilities of sound scenes are provided from ASC models.

Secondly, it is fact that sound events and sound scene in the same recording present a high correlation. 
For an instance, gun sound can be only detected in a riot context or a group of sound events such as wind, grass, and bird song is usually detected in a park.
Therefore, it is potential and feasible to use sound event detection (SED) results to support an ASC system.
Indeed, this has already been proven in Section~\ref{event_per} or some recent published papers~\cite{scene_by_event_01, jung_asc_aed}.
However, these work only focus on how to enhance the accuracy performance instead of using sound event information to present the a sound context more comprehensive.

These two reasons recently mentioned inspires us to propose a visualization method, which not only reports predicted probabilities of sound scene contexts, but it also visually presents a sound scene context more comprehensive by leveraging sound event information.

\subsection{Dataset and the use case definition}
\label{visual_dataset}
\begin{figure}[t]
    \centering
    \includegraphics[width =1.0\linewidth]{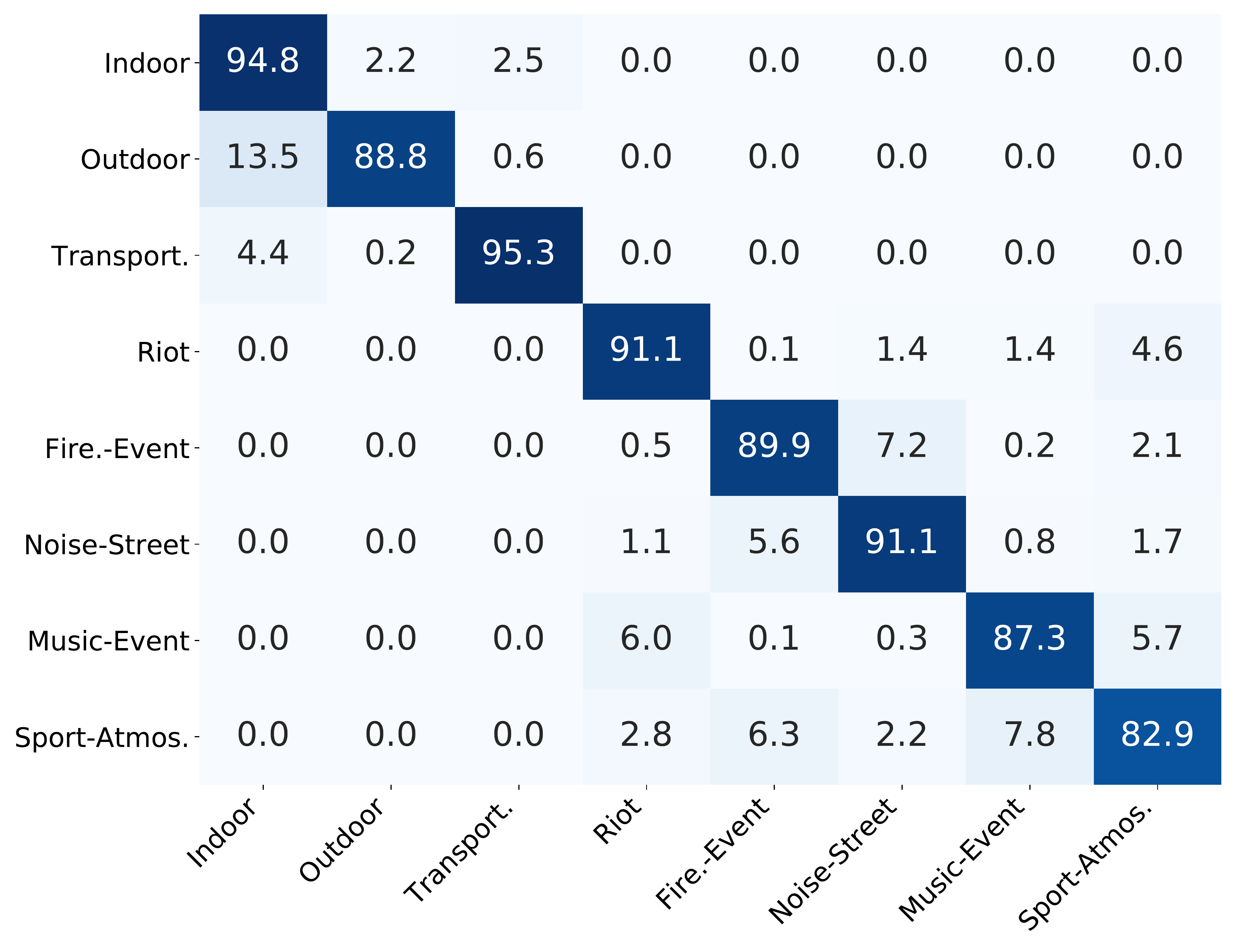}
     \vspace{-0.3cm}
	\caption{Confusion matrix results for ASC task on 8-sound-scene-context dataset}
    \label{fig:confuse}
\end{figure}

To evaluate the proposed visualization method, we firstly define a dataset and propose a case study for applying this method.
Regarding the evaluating dataset, we combine DCASE 2020 Task 1A Development and Crowded Scenes to form a new dataset of 15 sound scene contexts: Airport, Bus, Metro, Metro-Station, Park, Public-Square, Shopping-Mall, Street-Pedestrian, Street-Traffic, Tram, Music-Event, Sport-Event, Firework, Noise-Street, and Riot.
These 15 sound scenes are then grouped into 8 main categories: Daily Indoor (Airport, Shopping-Mall, and Metro-Station), Daily Outdoor (Park, Public-Square, Street-Pedestrian, and Street-Traffic), Daily Transportation (Bus, Metro, and Tram), Music-Event, Sport-Event, Firework, Noise-Street, and Riot, which is referred to as 8-sound-scene-context dataset.

Given the 8-sound-scene-context dataset, we define a specific task (e.g., the case study) which satisfies three requirements of: (1) detect a riot context from the 8-sound-scene-context dataset recently defined (i.e., In the other words, the requirement (1) is a task of sound scene classification on 8-sound-scene-context dataset), (2) low-complexity classification model with less than  5 M trainable parameters (approximately 20 MB memory occupation using 32-bit floating point to present 1 model trainable parameter) which is potential to integrate into a wide range of edge devices and mobiles, and (3) a visualization method for well presenting statistic information of sound events which matches a riot context.

Experimental settings for evaluating the proposed visualization method are reused from Section~\ref{baseline_dataset}, but using two datasets of DCASE 2020 Task 1A Development and Crowded Scenes.

\subsection{Propose an audio based system for detecting and presenting a riot context}
\label{visual_usecase}

To met the requirements of (1) and (2), we reuse the results from Section~\ref{event_section}.
In particular, we use two models of SPECs-NRI-RD64 and DS-ASC-MobV2 as shown in Table~\ref{table:event_res}, presenting 4.96 M of trainable parameters (19.4 MB memory occupation), to train and evaluate on the the 8-sound-scene-context dataset. 
While SPECs-NRI-RD64 is only for ASC task, DS-ASC-MobV2 is used not only for improving ASC task as experimental results in Section~\ref{event_per} but also for detecting sound events occurring in the sound context.
Given the result of ASC and sound events detected, we meet the requirement (3) by generating figures to describe the relationship between sound events and sound scene.
\begin{figure}[t]
    \centering
    \includegraphics[width =1.0\linewidth]{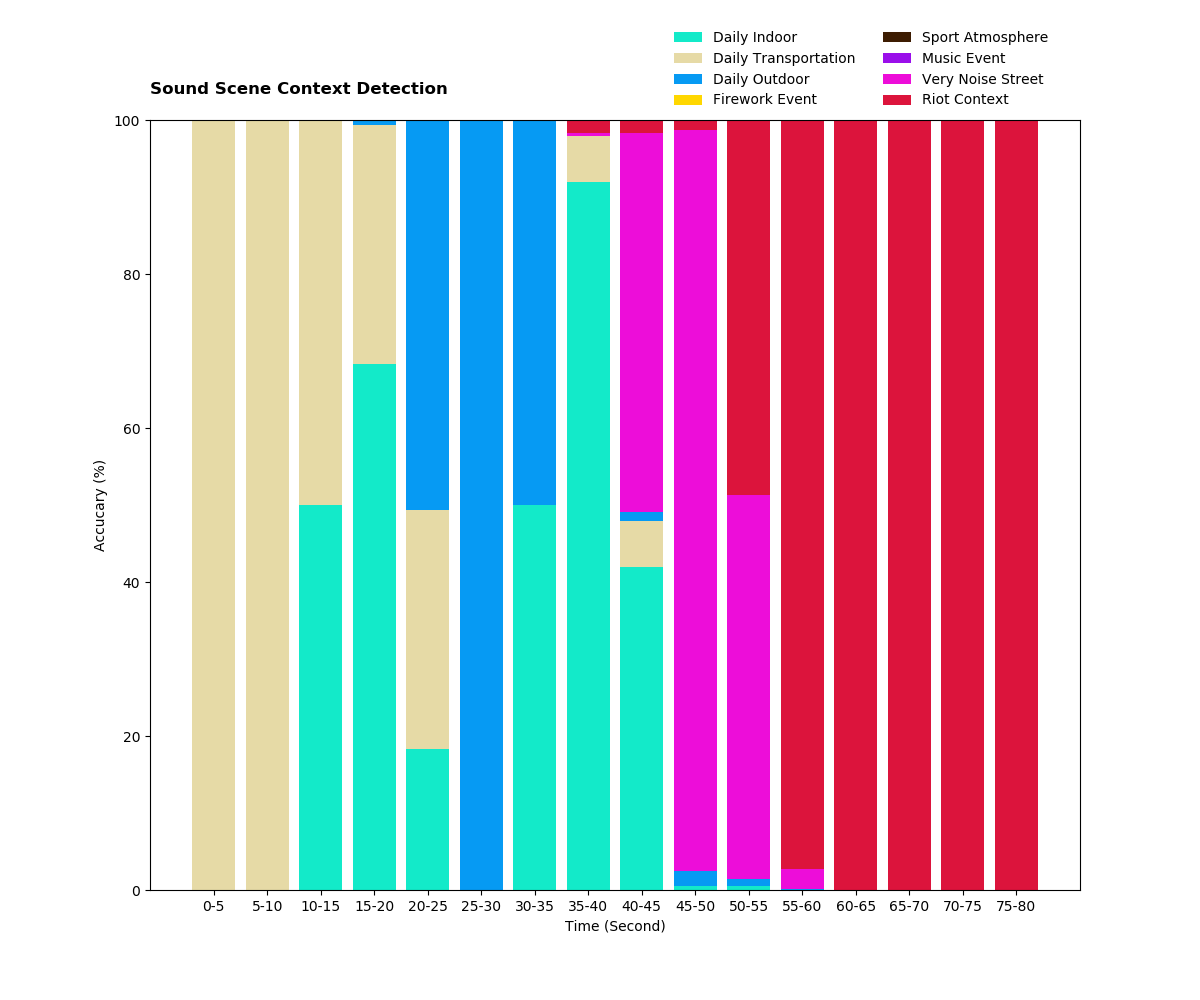}
     \vspace{-1cm}
	\caption{Visualization method: Presenting accuracy of detected sound scene contexts and transferring between two different contexts}
    \label{fig:vf1}
\end{figure}

To better describe relevant sound events which match riot contexts regarding the requirement (3), we propose two methods to separate sound events into certain groups. 
The first method is presented in Table~\ref{table:sound_event_level} which separates 527 types of sound events defined in AudioSet dataset into three main different alarming levels.
In particular, we have three levels, namely Red Level, Yellow Level, and Green Level as shown in Table~\ref{table:sound_event_level}.
The Red Level presents very dangerous and rare sound events which are only found in violent scenes.
The sound events with Yellow Level cause a negative or annoying felling which are separated into four sub-categories: sound events made by individual human, sound events from a crowd, natural sound events, and sound events from machines or things.
The other sound events are grouped into the Green Level, which is considered as normal events in daily life.
In the second method, we separate 527 types of sound events into 7 main categories: Human, Music, Things, Acoustic, Nature, Machine or Vehicle, and Animal.
While the first grouping method is driven from our statistics on sound events occurring in riot context, which is conducted on riot recordings in Crowded Scene dataset, the second method is based on the ontology of AudioSet dataset introduced by Google in~\cite{audioset_ont}.
These two methods help to describe how groups of sound events are relevant to a riot context.
\begin{figure}[t]
    \centering
    \vspace{-0.5cm}
    \includegraphics[width =1.0\linewidth]{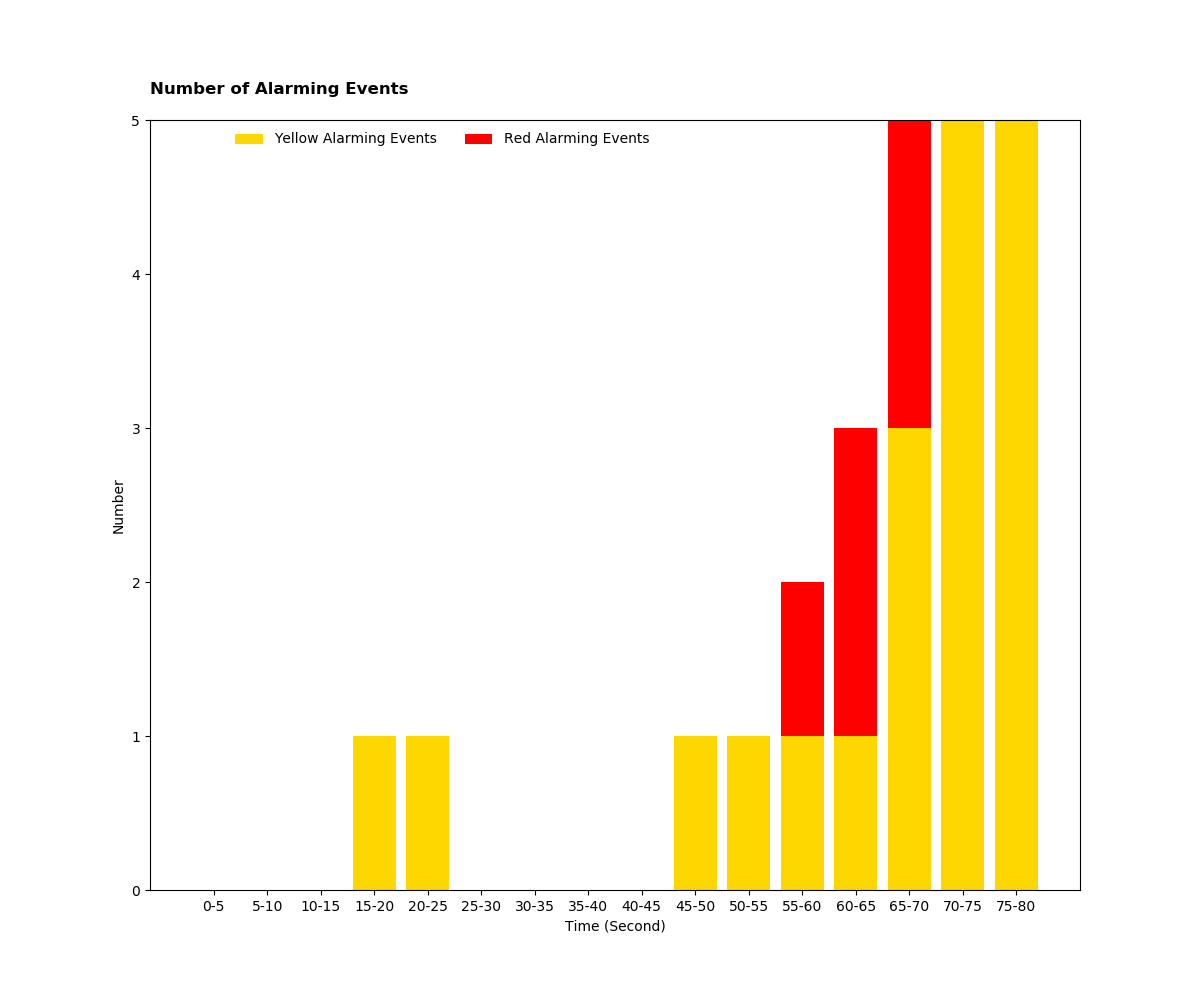}
     \vspace{-1cm}
	\caption{Visualization method: Presenting Red and Yellow alarming sound events numbers}
    \label{fig:vf2}
\end{figure}

\begin{figure}[t]
    \centering
    \vspace{-0.3cm}
    \includegraphics[width =1\linewidth]{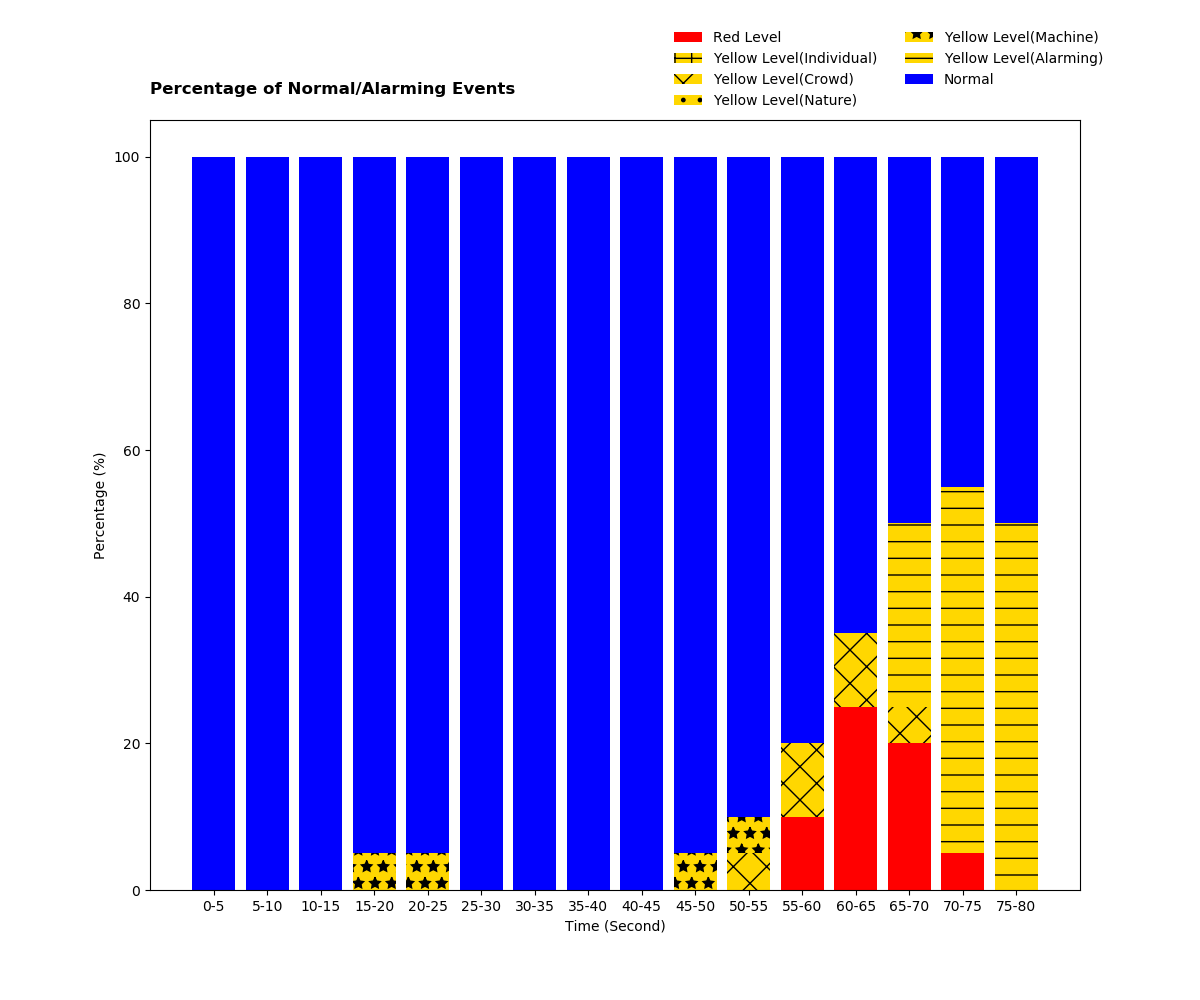}
     \vspace{-1cm}
	\caption{Visualization method: Presenting percentage of Red, Yellow, and Green sound events numbers on each 5-second segment}
    \label{fig:vf3}
\end{figure}

\begin{figure}[t]
    \centering
    \vspace{-0.5cm}
    \includegraphics[width =1\linewidth]{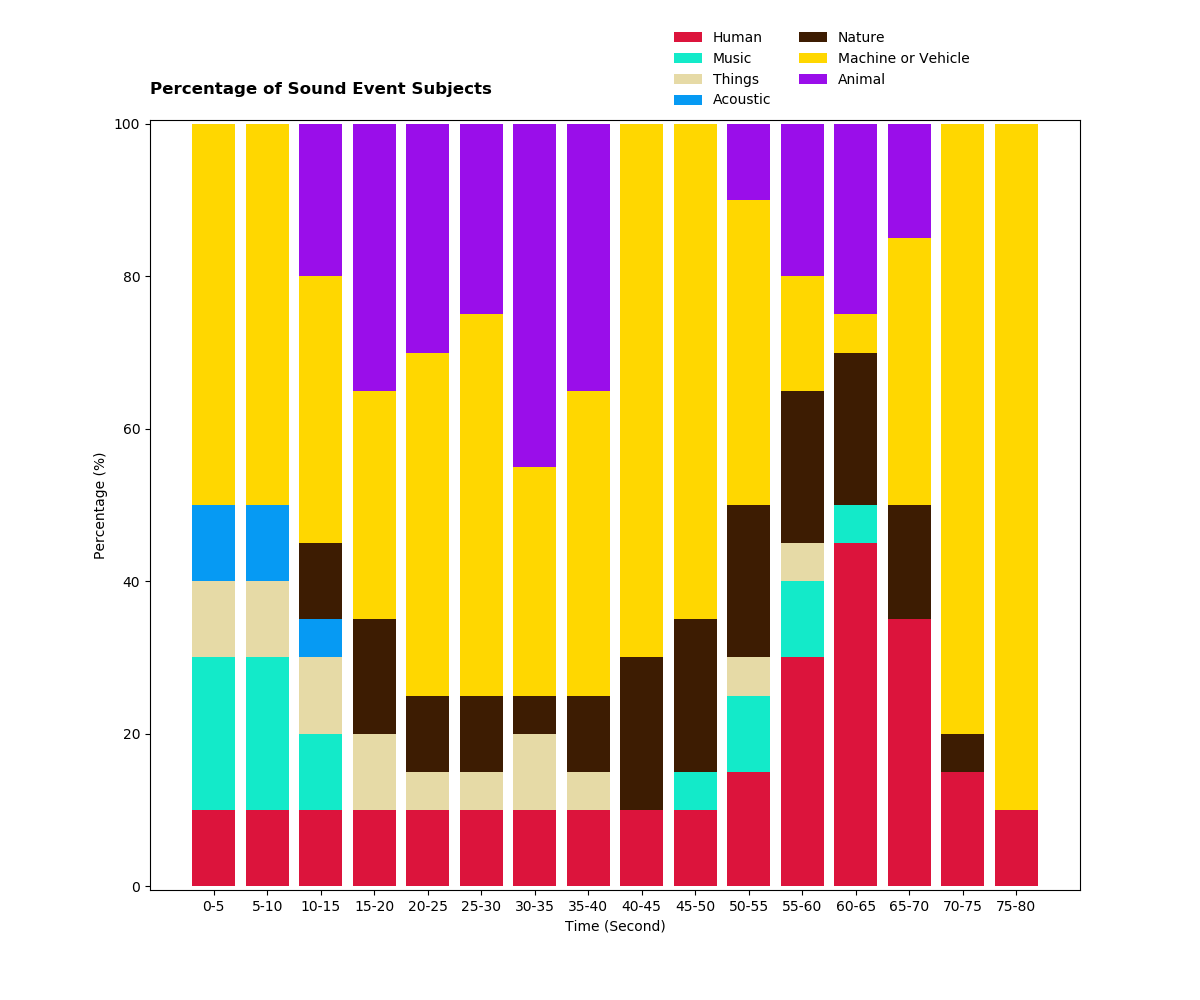}
     \vspace{-1cm}
	\caption{Visualization method: Presenting topology group of sound events on each 5-second segment}
    \label{fig:vf4}
\end{figure}
Overall, we expect that a sound scene of a riot context in the proposed case study can be presented more comprehensive by exploring both sound scene information (e.g., predicted probabilities of a sound recording) and sound event information (e.g., statistics and visualization of sound events in a sound recording).

\subsection{Experimental results}
\label{visual_res}

Figure~\ref{fig:confuse} presents a confusion matrix of 8 classes which is the classification result of SPECs-NRI-RD64 and DS-ASC-MobV2 on 8-sound-scene-context dataset.
As Figure~\ref{fig:confuse} shows, the accuracy on each class is larger than 80\% and the overall accuracy achieves 90.9\%.
As the proposed model (SPECs-NRI-RD64 and DS-ASC-MobV2) proves robust and presents a low footprint of 19.4 MB memory occupation on the 8-sound-scene-context dataset, the model is very potential to apply on various mobiles or edge devices.
\begin{table*}[t]
    \caption{Compare our proposed ASC systems (Large model (LM) with SPECs-NRI and DS-ASC-CNN14; Medium Model (MM) with SPECs-NRI-64 and DS-ASC-Mobv2, and Small Model (SM) with SPECs-NRI-120KB w/ quantization; Submitted models: Models submitted to DCASE 2021 and 2022 challenges) to the state-of-the-art systems on various sound scene datasets} 
        	\vspace{-0.2cm}
    \centering
    \scalebox{0.8}{

    \begin{tabular}{|lc| lc|  lc|  lc|    lc|  lc|  lc|  lc|} 
        \hline 
	        \textbf{DC-18-1A} &\textbf{Acc.\%}    &\textbf{DC-18-1B} &\textbf{Acc.\%}    &\textbf{DC-19-1A} &\textbf{Acc.\%}   &\textbf{DC-19-1B} &\textbf{Acc.\%}     &\textbf{DC-20-1A} &\textbf{Acc.\%}    &\textbf{DC-21-1A} &\textbf{Acc.\%}   &\textbf{DC-22} &\textbf{Acc.\%}  \\ 
	        
	       \textbf{(dev. set)} &    &\textbf{(dev. set)} &    &\textbf{(dev. set)} &   &\textbf{(dev. set)} &  &\textbf{(dev. set)} &    &\textbf{(test. set)} &   &\textbf{(test. set)} &   \\

        \hline 
	 Wang~\cite{wang_1819a}         &72.4          &DCASE baseline~\cite{}      &45.6          &DCASE baseline            &63.3           &DCASE baseline              &47.7           &Jung~\cite{jung_asc_aed} &70.4          &top-1    &\textbf{76.1}  &top-1     &\textbf{60.8}  \\
         Zhao~\cite{zhao_dca_18_ica}    &72.6          &Tchorz~\cite{tch_dca_18}    &53.9          &Wang~\cite{wang_1819a}    &75.7           &Wang~\cite{wan_dc_19b}      &55.2           &Shim~\cite{shim_cp_01}   &71.3          &top-2    &73.1           &top-2     &59.7           \\
         Zhao~\cite{zhao_dca_18}        &72.7          &Shefali~\cite{she_dca_18}   &56.2          &Jung~\cite{jun_2019_t06}  &76.2           &Jiang~\cite{jia_dc_19b}     &64.2           &Zhao~\cite{adap01}       &72.2          &top-3    &72.1           &top-3     &56.3           \\
         Phaye~\cite{phaye_dca_18}      &74.1          &Zhao~\cite{Zhao_cp01}       &63.3          &Javier~\cite{javier_cp01} &76.7           &Primus~\cite{prim_dc_19b}   &65.1           &Choi~\cite{choi_cp_01}   &72.3          &top-4    &71.8           &top-4     &55.2           \\
         Jung~\cite{jee_dnn}            &74.8          &Truc~\cite{truc_dca_18}     &63.6          &Cho~\cite{cho_cp01}       &77.2           &McDonnell~\cite{don_dc_19b} &66.3           &Liu~\cite{t5_dc_20a}     &73.1          &top-5    &70.1           &top-5     &54.9           \\
         Hossein~\cite{hos_dca_18}      &76.8          &Truc~\cite{truc_dca_18_int} &64.7          &Mars~\cite{roh_dc_18}     &79.3           &Zhao~\cite{adap01}          &66.5           &Koutini~\cite{t4_dc_20a} &73.3          &top-6    &69.6           &top-6     &54.7           \\
         Heo~\cite{Heo}                 &77.4          &Truc~\cite{truc_dca_18_icme}&66.1          &Choi~\cite{choi_cp_01}    &81.1           &Song~\cite{song_dc_19b}     &70.3           &Suh~\cite{t3_dc_20a}     &74.2          &top-7    &68.8           &top-7     &53.8           \\
         Hou~\cite{hou_cp01}            &77.4          &Yang~\cite{yang_cp01}       &67.8          &Wang~\cite{wang_cp01}     &82.6           &Michal~\cite{kos_dc_19}     &74.0           &Ma~\cite{ens_06}      &75.0          &top-8    &68.5           &top-8     &52.7           \\
	 Octave~\cite{octave_exploring} &79.3          &Wang~\cite{jun_dca_18}      &69.0          &Huang~\cite{hua_2019_t05} &83.1           &                            &               &Wang~\cite{wan_dc_20a}   &\textbf{81.8} &top-9    &68.3           &top-9     &52.7           \\
         Yang~\cite{yang_acoustic}      &\textbf{79.8} &Lam~\cite{lam_jour}         &70.6          &Koutini~\cite{kou_dc_19}  &\textbf{83.7}  &                            &               &                         &              &top-10   &68.1           &top-10    &51.7           \\
        \hline      
        \hline      
        Our LM                         &79.3           &Our LM                      &\textbf{73.3} &Our LM                    &81.3           &Our LM                      &\textbf{75.1}   &Our LM                   &75.4          &Submitted &69.6          &Submitted &55.2          \\
	Our MM                         &77.8           &Our MM                      &\textbf{73.0} &Our MM                    &78.5           &Our MM                      &\textbf{70.5}   &Our MM                   &73.9          &Model     &              &Model     &              \\
	Our SM                         &71.6           &Our SM                      &57.2          &Our SM                    &73.5           &Our SM                      &66.6            &Our SM                   &73.9          &          &              &          &              \\
        \hline 

    \end{tabular}
    }
    \label{table:sota_res} 
\end{table*}

To present how results of sound scene classification and statistic information of sound events are well presented via the proposed visualization method, we set up an 80-second recording which resents different sound scene contexts: `in metro' from 0 second to 10 seconds, `in metro station' from 10 seconds to 20 seconds, 'in traffic street' from 20 seconds to 30 seconds, 'in shopping mall' from 30 seconds to 40 seconds, 'in very noise street' from 40 seconds to 50 seconds, and finally 'in a riot context' from 50 second to 80 seconds.
Given the audio recording, we fed into the proposed the system (SPECs-NRI-RD64 and DS-ASC-MobV2), present results as shown Figure~\ref{fig:vf1},~\ref{fig:vf2},~\ref{fig:vf3}, and~\ref{fig:vf4}.
We can see that Figure~\ref{fig:vf1} presents sound scenes detected on each 5-second segment.
When the sound scene changes at a certain time (e.g., `in metro' to `in metro station' at the time slot of 10 seconds), Figure~\ref{fig:vf1} shows both sound scenes before and after this time point.
As the proposed case study aims at detecting whether a riot context occurs in an audio recording, the riot context is detected and marked with the red color as shown from 50 seconds to 80 seconds in Figure~\ref{fig:vf1}.

As the riot context is detected from 50 second to 80 second, we check groups of alarming sound events recently proposed in~\ref{visual_usecase}.
As Figure~\ref{fig:vf2} and~\ref{fig:vf3} show, we can see that both Red Level alarming sound events and Yellow Level alarming events are detected from 50 seconds to 80 seconds.
While Figure~\ref{fig:vf2} presents the number of alarming events, Figure~\ref{fig:vf3} shows the ratios of these alarming sound events on each 5-second segment.
We also see that the `Alarming' sub-group of Yellow Level events mainly appear in the riot context detected from 50 seconds to 80 seconds.
The final Figure~\ref{fig:vf4} indicates that almost sound events related to machines/devices or human occur in a riot context.

Our above experiments have proven that a riot context can be indicated and comprehensive analyzed on a wide range of devices or mobiles by using models of SPECs-NRI-RD64, DS-ASC-MobV2 and the proposed visualization method.
By early detect certain riot contexts, it helps to predict a possible large-scale migration or trigger immediately a warning for a certain region (e.g., a violent riot is occurring at the street/district/country X) before the mainstream media (e.g., Television channels, newspaper, etc.) reports. 
Given the comprehensive analysis of our case study, an application of detecting and presenting a sound scene context can be feasibly developed and implemented on a wide range of edge devices and mobiles. 

\section{Compare our proposed ASC systems to the state-of-the-art systems}
\label{sota_section}

To compare with state-of-the-art ASC systems, we propose three models in this paper: large-size model (LM) which combines SPECs-NRI and DS-ASC-CNN14 (i.e., DS-ASC-CNN14 is the down-stream ASC task using the up-stream pre-trained CNN14 model in~\cite{pann_paper}); medium-size model (MM) which combines SPECs-NRI-RD64 and DS-ASCMobv2 presenting 4.96 M of trainable parameter and occupying 19.4 MB memory (i.e., This model was evaluated in the upper sections of~\ref{visual_section} and~\ref{event_section}); small-size model (SM) which used SPECs-NRI-120KB with quantization presenting 0.12 M of trainable parameters and occupying 120 KB memory (i.e., This model was evaluated in the upper Section~\ref{tradeoff_section}).
All three models are evaluated on a wide range of ASC datasets mentioned in Section~\ref{data_set}: DCASE 2018 Task 1A, DCASE 2018 Task 1B, DCASE 2019 Task 1A, DCASE 2019 Task 1B, DCASE 2020 Task 1A.
For DCASE 2021 Task 1A and DCASE 2020 Task 1, we report the results which are from our submitted models presented in~\cite{pham2021low} and~\cite{pham2022low}, respectively.
Notably, the submitted models also make use of CNN-based network architecture and model compression techniques of channel reduction (CR), channel deconvolution (CD), and Quantization (Qu.).

As results are shown in Table~\ref{table:sota_res}, we can see that our large-size model (LM) outperform state-of-the-art ASC systems in DCASE 2018 Task 1B, DCASE 2019 Task 1B, achieves the top-2 in both DCASE 2018 Task 1A and DCASE 2020 Task 1A, and occupies the top-4 in DCASE 2019 Task 1A.
Note that these tasks do not require the low-complexity model and almost state-of-the-art systems made use of high-complexity models and ensemble methods.

On these datasets, although the accuracy performance of our proposed medium-size model (MM) slightly reduces as this model is constrained by maximum 5 M of trainable parameter and occupying 20 MB memory to be able to apply on a wide range of edge devices and mobiles, the results are still very competitive to the state-of-the-art systems (top-3 in DCASE 2018 Task 1A, top-1 in DCASE 2018 Task 1B, top-6 in DCASE 2019 Task 1A, top-2 in DCASE 2019 Task 1B, and top-4 in DCASE 2020 Task 1A).

Regarding our proposed Small Model (SM) which is constrained by maximum 128 KB memory occupation, this model is still in top-10 compared to state-of-the-art systems on all evaluating datasets. 
Specially, this model achieves top-3 and top-4 in DCASE 2019 Task 1B and DCASE 2020 Task 1A. 
Our submitted systems for DCASE 2021 Task 1A and DCASE 2022 Task 1 (These challenges requires low complexity model with the same constrain of maximum 128 KB memory occupation) also achieve top-6 and top-4 rankings, respectively.

\section{Conclusion}
\label{conclusion}
This paper has presented robust, general, and low complexity systems for acoustic scene classification (ASC), achieved three main outcomes.
Firstly, by using multiple techniques: a novel inception-residual based network architecture, an ensemble of multiple spectrogram inputs, ASC down-stream task inherited from the up-stream SED task, and model compression methods, we achieve very competitive ASC systems compared to the state of the art on almost currently challenging ASC datasets.
Secondly, we propose two low-complexity network ASC models: medium-size model (19.4 MB memory occupation) and small-size model (128 KB memory occupation), which are suitable for real-life applications on a wide range of edge devices and mobiles. 
Finally, a visualization method has been proposed to present a sound scene context more comprehensive by exploring both sound event and sound scene information.





\bibliographystyle{IEEEbib}
\bibliography{refs}


\end{document}